\documentclass{aa}
\usepackage{graphicx}
\usepackage{longtable}
\usepackage{latexsym}
\usepackage{amssymb}
\usepackage{lscape}
\input{psfig.txt}
\begin{document}
\title{
Molecular gas in normal late-type galaxies
\thanks{based on observations made with the 12-m National Radio Astronomical
Observatory, Kitt Peak, Arizona}}

\author{A. Boselli \inst{1}
\and J. Lequeux \inst{2}
\and G. Gavazzi \inst{3}
}

\authorrunning{A.Boselli et al.}
\titlerunning{Molecular gas in normal late-type galaxies}

\offprints{Alessandro Boselli}

\institute{
Laboratoire d'Astrophysique de Marseille, BP 8, 
Traverse du Siphon, F-13376 Marseille Cedex 12, France\\
\email {Alessandro.Boselli@astrsp-mrs.fr}
\and DEMIRM and URA 336 du CNRS, Observatoire de Paris, 61 Av. de 
l'Observatoire, F-75014 Paris, France \\
\email {James.Lequeux@obspm.fr}
\and Universit\`a degli Studi di Milano-Bicocca, Dipartimento di Fisica, 
Piazza dell'Ateneo Nuovo 1, 20126 Milano, Italy\\
\email {Giuseppe.Gavazzi@mib.infn.it}
}

\date{}

\abstract{We present $^{12}$CO(J=1--0) line observations of 22 low-luminosity spiral galaxies 
in the Virgo cluster. These data, together with 244 others 
available in the literature,
allow us to build a large sample that we use to study the molecular gas 
properties of galaxies spanning a large range of morphological types
and luminosities and belonging to different environments (clusters - field).
The molecular gas content of the target galaxies is estimated using
a luminosity-dependent $X$ = $N(H_2)$/$I(CO)$ conversion factor that has 
been calibrated on a sample of nearby galaxies. 
$X$ spans from $\sim$ 10$^{20}$ mol cm$^{-2}$ (K km s$^{-1}$)$^{-1}$ in giant
spirals to $\sim$  10$^{21}$ mol cm$^{-2}$ (K km s$^{-1}$)$^{-1}$ in dwarf irregulars. 
The value of the $X$ conversion factor is found consistent with a value derived 
independently from dust masses estimated from FIR fluxes, with a metallicity-dependent 
dust to gas ratio.
The relationships between $X$ and the UV radiation field
(as traced by the $H\alpha+[NII] E.W.$), the metallicity and the H band
luminosity are analysed. 
We show that the molecular gas contained in molecular clouds or complexes
is of the order of 15\% of the total gas on average whatever the luminosity
or the Hubble type of the galaxies. We discuss the relation between the star formation 
rate and the molecular gas content and estimate the average star formation efficiency
of late-type galaxies.
}

\maketitle
%

\section{Introduction}

Our understanding of galaxy evolution is still limited by the poor
knowledge of the physical mechanisms behind the process of star formation, i.e.
the condensation of the primordial gaseous material (HI) into
molecular clouds (H$_2$), then their collapse into stars. Also poorly understood are
the feed-back processes induced by the newly born stars on the interstellar medium (ISM). 
The UV radiation field associated with massive star formation 
produces the ionization of the surrounding gas and causes the photo-dissociation
of the hydrogen molecules. Meanwhile supernova winds inject kinetic energy (and metals) into
the ISM with the effect of triggering new star formation on the
shock fronts. \\
The first and last stages of this complex chain of transformations
are constrained observationally: 
the atomic hydrogen is directly and easily observable through 
the 21 cm line, and the star formation rate (SFR) can be estimated
from H$\alpha$ and UV luminosities, once corrected for extinction,
using stellar population synthesis models (Kennicutt 1998).\\
Kennicutt (1989) suggested the existence of a
universal relationship between gas surface density at disc scales and
star formation activity, known as the Schmidt law (Schmidt 1959),
modulated by differential rotation.
Even if the total amount of gas in late-type galaxies is dominated by
atomic hydrogen (Boselli et al. 1997b), which commonly spans
a diameter $\sim$ 1.8 times the optical disc (Cayatte et al. 1994), 
the molecular phase might dominate the gas column density inside
the optical disc (Young \& Scoville 1991). 
An accurate determination of the content and distribution of 
the molecular gas is thus crucial for studying the interplay between
gas and star formation over galactic discs.\\
The amount of molecular hydrogen, can only be estimated by indirect methods.
The most common method starts from the $^{12}$CO(1-0) line emission
at 115 GHz using a $standard$ $X$ conversion factor 
between the intensity of the CO line ($I(CO)$) and the
column density of H$_2$ ($N(H_2)$) calibrated
in the solar neighborhood, $X$=$N(H_2)$/$I(CO)$ (Young \& Scoville 1991).
Detailed analyses of giant molecular clouds (GMC) in nearby galaxies
have however questioned the validity of
this method. $X$ can vary by up to a factor of $\sim$ 10
as observed in giant molecular clouds in the SMC (Rubio et al. 1993) 
or in M31 (Allen \& Lequeux 1993).
Higher values of $X$ are generally observed at 
low metallicities and strong UV radiation fields. \\
Using the few local galaxies for which $X$ has been determined with
independent methods, it might be possible to calibrate some
empirical relationships between $X$ and other parameters characterizing
the ISM, such as metallicity,
intensity of the UV radiation field and cosmic ray density.
This approach is however not easy since the determination of
$X$ relies on molecular gas mass estimates based on different techniques:
from the virial equilibrium of GMCs (Young \& Scoville 1991), from the 
line ratios of different CO isotopes (Wild et al. 1992), assuming
a  metallicity-dependent gas to dust ratio (Gu\'elin et al. 1993, 1995), 
or relying on $\gamma$-ray data (Hunter et al. 1997).
\noindent
Furthermore the relation between the CO emission properties and 
the physical parameters of the ISM (UV radiation field, cosmic ray density,
metallicity, gas density) are still poorly known, even though some
success in modeling the ISM has recently been reported (Lequeux et al. 1994, 
Kaufman et al. 1999, Bolatto et al. 1999).\\
An ideal approach in determining $X$ would be to use a consistent method 
on a sample of galaxies spanning a large range in luminosity
and morphological type. The technique applied by Gu\'elin and collaborators 
to a few nearby galaxies (Gu\'elin et al. 1993, 1995, Neininger 
et al. 1996; see also Israel 1997) seems very promising. It consists of
determining the amount of dust from millimetric/submillimetric observations
and correlating it with CO and HI. 
Provided that the gas to dust ratio is constant 
or metallicity-dependent in a known way, the molecular gas mass can be determined
assuming that the difference between the expected total gas to dust ratio
and the observed HI to dust ratio is due to molecular gas.
This technique could be profitably applied to larger galaxy samples provided that 
the necessary multifrequency data are available.\\
For the last few years we have been gathering spectrophotometric data
from the UV to the radio centimetric domain for
a large sample of $\sim$ 2500 galaxies in the nearby Universe. 
The database created so far is ideal for this purpose since it includes
the data necessary for the determination of $X$ (HI, CO and far-IR fluxes)
as well as those needed to characterize the physical properties of
the ISM (H$\alpha$ fluxes as tracers of the UV radiation field 
and metallicity measurements). 
The present dataset is however still limited
to the bright end of the luminosity function. In particular we are 
lacking CO and metallicity measurements for galaxies with $M_B$ $>$ -18.\\
In order to extend the available CO data to lower luminosities, we observed  
22 galaxies with -19$<$ $M_B$ $<$-16 in the Virgo cluster. 
The new observations are described in sect. (2, 3, 4). These new data
are combined in sect. (5) with the many data points 
already available for the galaxies 
included in our database in order to derive $X$ and then their molecular content.
The results of our analysis are discussed and compared 
with those obtained for a sample of 14 nearby galaxies with independent
measurements of $X$ (sect. 6). 
The agreement in the determination of $X$ allows
us to calibrate a luminosity-dependent $X$ conversion factor
which is later used to re-analyse the molecular gas statistical properties 
of late-type galaxies (sect. 7). 
After correcting for the systematic trend of $X$ with luminosity,
we estimate the molecular gas fraction and the star formation efficiency of galaxies 
spanning a large range in luminosity and morphological type and belonging 
to different environments (sect. 7).  

\section {The new sample}

We report on new observations of the 
$^{12}$CO(1--0) line emission of 22 galaxies belonging to the Virgo Cluster. 
These are mostly low luminosity late-type galaxies.

\noindent
The target galaxies are listed in Table 1, arranged as follows:

\begin {itemize}
\item {Column 1: VCC designation (Binggeli et al. 1985). }
\item {Column 2 and 3: NGC/IC and UGC names.}
\item {Columns 4 and 5: 1950 celestial coordinates with a few 
arcsec accuracy, from Binggeli et al. (1985).}
\item {Column 6: Morphological type, 
from Binggeli et al. (1985).}
\item {Column 7: Membership of the Virgo cluster, according to Gavazzi et al. (1999b).} 
\item {Column 8: Heliocentric velocity, in km s$^{-1}$, from Binggeli 
et al. (1985).}
\item {Column 9: Distance in Mpc, determined from the cluster membership
as described in Gavazzi et al. (1999b).}
\item {Columns 10 and 11:  Galaxy major and minor optical blue diameters, in 
arcminutes, from Binggeli et al. (1985). These are isophotal diameters at 
the faintest observable magnitude.} 
\item {Column 12: Photographic magnitude as given in the VCC.}
\item {Column 13: Width of the HI line (in km s$^{-1}$) calculated 
averaging the width at 20\% and at 50\% of the maximum intensity, from 
Helou et al. (1984), Hoffman et al. (1987; 1989), 
Schneider et al. (1990), Haynes \& Giovanelli (1986), Magri (1994).}
\item {Column 14: Total extrapolated H magnitude, from Boselli et al. 
(1997a; 2000)}
\end {itemize}

\begin{table*}

\caption{The target galaxies}
\label{Tab1}
\[
\begin{array}{rccllclrrlllcr}
\hline
\noalign{\smallskip}
VCC  & NGC/IC & UGC & RA(1950) & dec(1950) & type & agg & vel & dist & a & b & m_{pg} & \Delta V_{HI} & H_T\\ 
     &        &     &\rm{~ h ~m ~ s }&\rm{~^o ~' ~"} & & &\rm{ km~s^{-1}} &\rm{Mpc} &\rm{'} &\rm{'} &  &\rm{km~s^{-1}} & \\
\noalign{\smallskip}
\hline
\noalign{\smallskip}
     58& IC769&  7209& 120959.20& 122407.0 &  Sb& M&  2207 & 32.0 & 2.54 & 1.75 &13.17 &  252 & 10.51\\
     87&    - &    - & 121108.20& 154354.0 &  Sm& N&  -134 & 17.0 & 1.45 & 0.72 &15.00 &  109 & 13.64\\
     92&  4192&  7231& 121115.50& 151042.0 &  Sb& N&  -135 & 17.0 & 9.78 & 2.60 &10.92 &  469 &  7.07\\
     97&  4193&  7234& 121120.80& 132703.0 &  Sc& M&  2476 & 32.0 & 1.96 & 0.97 &13.20 &  367 &  9.61\\
    199&  4224&  7292& 121400.50& 074424.0 &  Sa& W&  2594 & 32.0 & 2.92 & 1.00 &12.95 &  554 &  8.87\\
    318& IC776&  7352& 121630.50& 090802.0 & Scd& W&  2469 & 32.0 & 1.71 & 1.00 &14.01 &  178 & 12.95\\
    459&    - &    - & 121839.60& 175457.0 & BCD& A&  2108 & 17.0 & 0.84 & 0.36 &14.95 &  127 & 12.70\\
    792&  4380&  7503& 122249.60& 101738.0 & Sab& B&   971 & 23.0 & 3.52 & 1.75 &12.36 &  290 &  8.55\\
    874&  4405&  7529& 122335.50& 162728.0 &  Sc& A&  1738 & 17.0 & 1.89 & 1.11 &12.99 &  169 &  9.61\\
    939&    - &  7546& 122414.70& 090940.0 &  Sc& B&  1271 & 23.0 & 3.45 & 3.45 &12.92 &   91 & 10.53\\
    957&  4420&  7549& 122425.20& 024618.0 &  Sc& S&  1695 & 17.0 & 2.01 & 0.85 &12.67 &  214 &  9.87\\
   1205&  4470&  7627& 122705.30& 080559.0 &  Sc& S&  2339 & 17.0 & 1.84 & 1.15 &13.04 &  147 & 10.22\\
   1290&  4480&  7647& 122753.40& 043127.0 &  Sb& S&  2438 & 17.0 & 2.01 & 1.07 &13.09 &  319 &  9.80\\
   1375&    - &  7668& 122906.20& 041256.0 &  Sc& S&  1732 & 17.0 & 4.76 & 3.77 &12.00 &  173 & 11.17\\
   1412&  4503&  7680& 122934.20& 112708.0 &  Sa& A&  1342 & 17.0 & 4.33 & 1.71 &12.12 &   -  &  8.28\\
   1508&  4519&  7709& 123057.90& 085549.0 &  Sc& S&  1212 & 17.0 & 3.60 & 2.60 &12.34 &  187 &  9.69\\
   1554&  4532&  7726& 123146.70& 064439.0 &  Sm& S&  2021 & 17.0 & 2.60 & 1.00 &12.30 &  185 &  9.76\\
   1686&IC3583&  7784& 123412.30& 133202.0 &  Sm& A&  1122 & 17.0 & 2.79 & 1.71 &13.95 &  116 & 11.19\\
   1929&  4633&  7874& 124006.50& 143748.0 & Scd& E&   291 & 17.0 & 2.48 & 1.07 &13.77 &  190 & 10.75\\
   1943&  4639&  7884& 124021.50& 133152.0 &  Sb& E&  1048 & 17.0 & 3.20 & 2.01 &12.19 &  295 &  8.90\\
   2023&IC3742&  7932& 124300.90& 133615.0 &  Sc& E&   958 & 17.0 & 2.01 & 1.00 &13.86 &  186 & 11.57\\
    -  &  4866&  8102& 125657.70& 142626.0 &  Sa& E&  1986 & 17.0 & 6.00 & 1.44 &11.90 &  537 &  8.19\\
\noalign{\smallskip}
\hline
\end{array}
\]
\end{table*}

\section {The observations} 

The observations were carried out during a remote-observing run (120 hours) from 
the Laboratoire d'Astrophysique de Marseille in June 2000 
using the NRAO Kitt Peak 12 m telescope
\footnote{The Kitt Peak 12-m telescope was operated by Associated Universities, 
Inc., under cooperative agreement with the National Science 
Foundation.}.
At 115 GHz [$^{12}$CO(1--0)], the telescope 
half-power beam width (HPBW) is 55" which corresponds to 4.5 kpc 
at the assumed distance of 17 Mpc for the Virgo cluster. 
Weather conditions were fairly good, with typical 
zenith opacities of 0.30-0.45. 
The pointing accuracy was checked every 
night by broad band continuum observations of Saturn and/or 3C273, with an 
average error of 7" rms. We used a dual-polarization SIS mixer, with a 
receiver temperature for each polarisation of about $T_{sys}$=350-600 K 
(in $T^*_R$
scale) at the elevation of the sources. We used a dual beam-switching 
procedure, with two symmetric reference positions offset by 4' in 
azimuth. The backend was a 256 channel filter bank spectrometer with channel 
width of 2 MHz. Each 6-minute scan began by a chopper wheel 
calibration on a load at ambient temperature, with a chopper wheel 
calibration on a cold load every two scans. Galaxies were observed 
at their nominal coordinates listed in Table 1, with one position per 
galaxy. The total integration time was on average 120 minutes on+off (i.e. 60 
minutes on the source), yielding rms noise levels of about 3 mK (in the 
$T^*_R$ scale) after velocity smoothing to 21 km s$^{-1}$. The baselines 
were flat owing to the use of beam-switching, thereby requiring that only 
linear baselines be subtracted. 
The antenna temperature $T^*_R$ was corrected for telescope and 
atmospheric losses. In the following analysis we use the main-beam 
brightness temperature scale, $T_{mb}$, with $T_{mb}$=$T^*_R$/0.84 
(where the main beam efficiency is $\eta$$_{mb}$=0.54 and the forward 
scattering and spillover efficiency $\eta$$_{fss}$=0.68).
This scale is 
appropriate for sources with sizes comparable to, or smaller than the 
beam size. These main-beam temperatures can be converted into 
flux densities using 29 Jy/K.

\section {Results}

\subsection{Results of our observations}

The $^{12}$CO(1--0) spectra of all the detected galaxies, reduced with the 
CLASS package (Forveille et al. 1990), are shown in Fig. 1: the 
observational results are listed in Table 2. Of the 22 observed galaxies 11 
were not detected. Table 2 is arranged as 
follows:

\begin{itemize}

\item {Column 1: VCC name.} 
\item {Column 2: Integration time (on+off), in minutes.}
\item {Column 3: rms noise, in mK, in the $T^*_R$ scale.}
\item {Column 4: Intensity of the $I(CO)$ line 
($I(CO)$=$\int$$T^*_R dv$) in K km s$^{-1}$ (area 
definition)). For undetected galaxies, the reported value 
is an upper limit determined as follows: 

\begin{equation}
{I(CO)=2 \sigma (\Delta V_{HI} \delta V_{CO})^{1/2}	 {\rm K~km~s^{-1}}}
\end{equation}

where $\sigma$ is the rms noise of the spectrum, $\Delta V_{HI}$ is the HI 
line width, and 
$\delta V_{CO}$ is the spectral resolution (for galaxies with 
$\Delta V_{HI}$ not available, the 
HI width has been determined assuming a standard $\Delta V_{HI}$ = 300 sin($i$) 
km s$^{-1}$, where $i$ is the galaxy inclination or 
$\Delta V_{HI}$ = 50 km s$^{-1}$ if $i$ = 0). 
$\delta V_{CO}$=21 km s$^{-1}$.}
\item {Column 5: Error on the intensity of the CO line, $\Delta$$I(CO)$, 
computed as: 

\clearpage
\begin{figure*}[!h]
\centerline{
\includegraphics[width=5.5cm,angle=-90,viewport=1 120 500 700,clip]{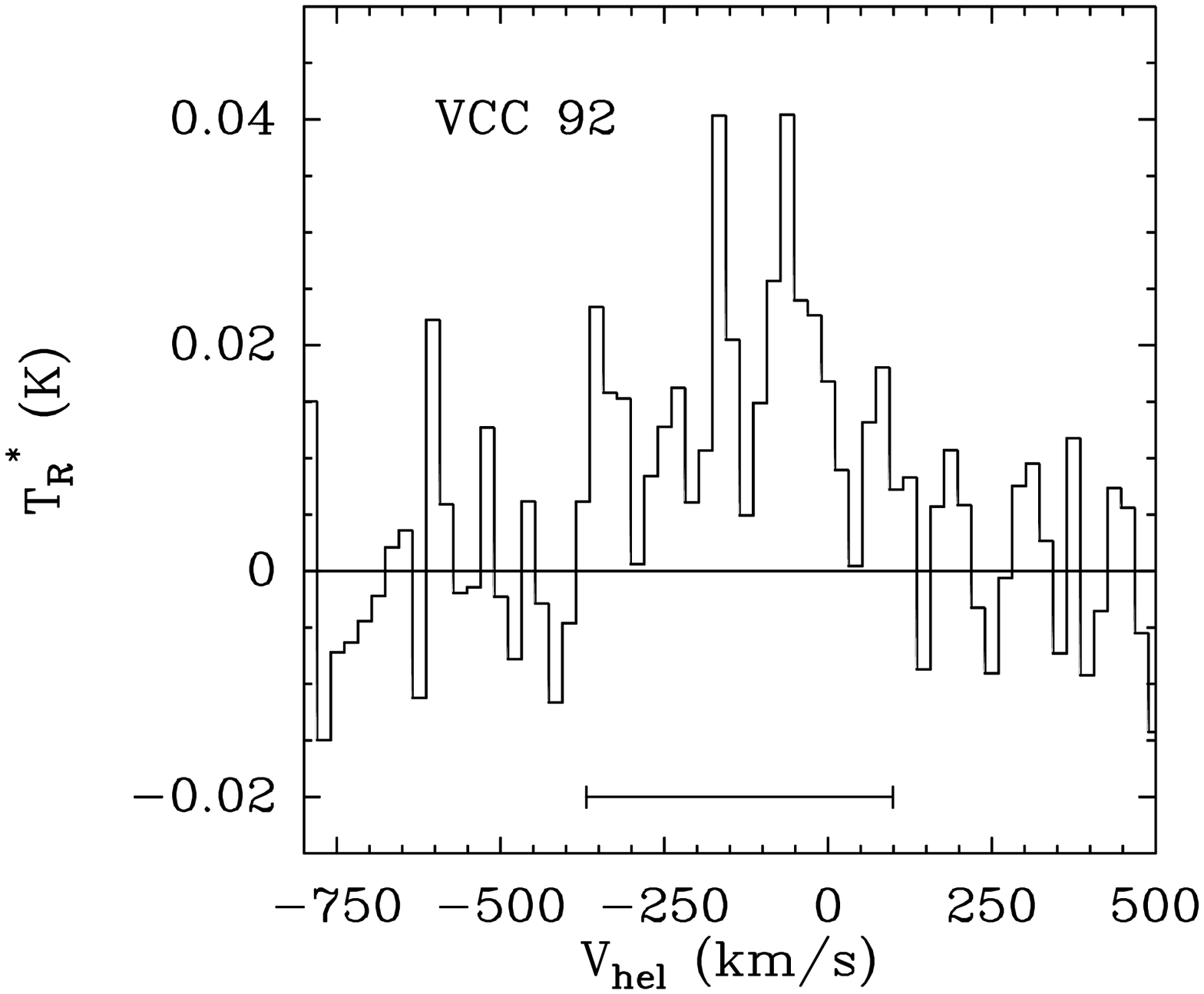}
\includegraphics[width=5.5cm,angle=-90,viewport=1 120 500 700,clip]{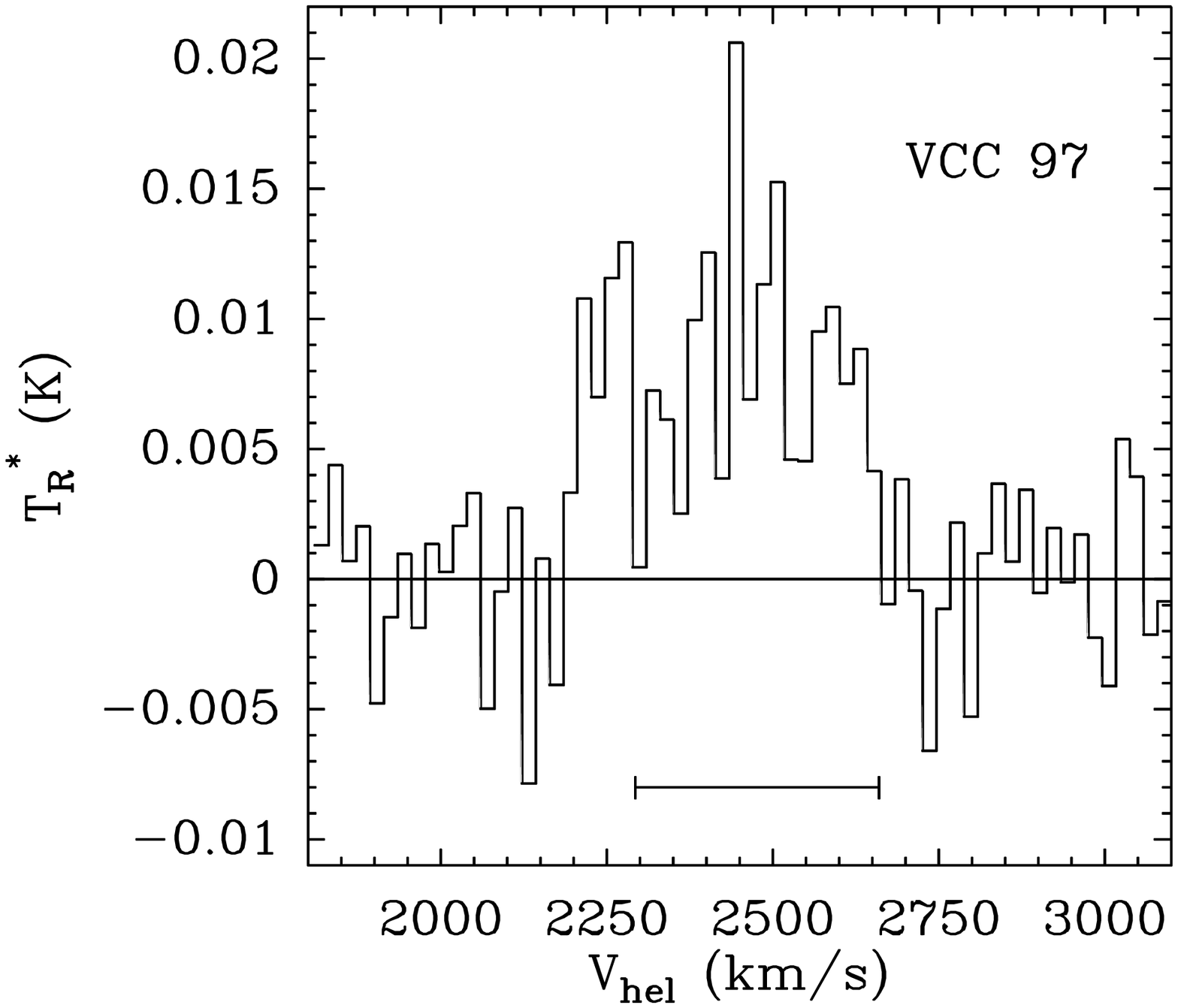}
\includegraphics[width=5.5cm,angle=-90,viewport=1 120 500 700,clip]{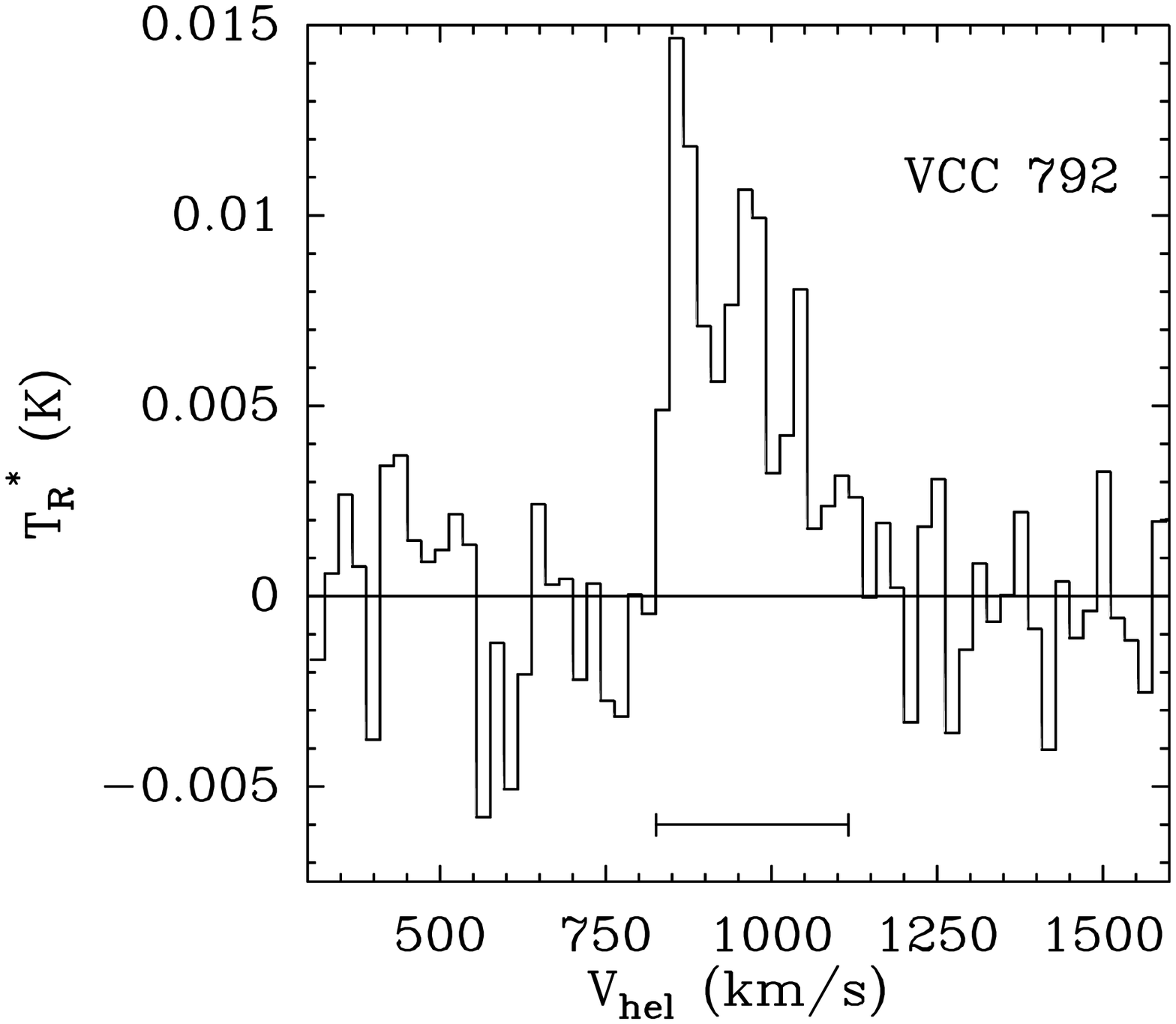}}
\centerline{
\includegraphics[width=5.5cm,angle=-90,viewport=1 120 500 700,clip]{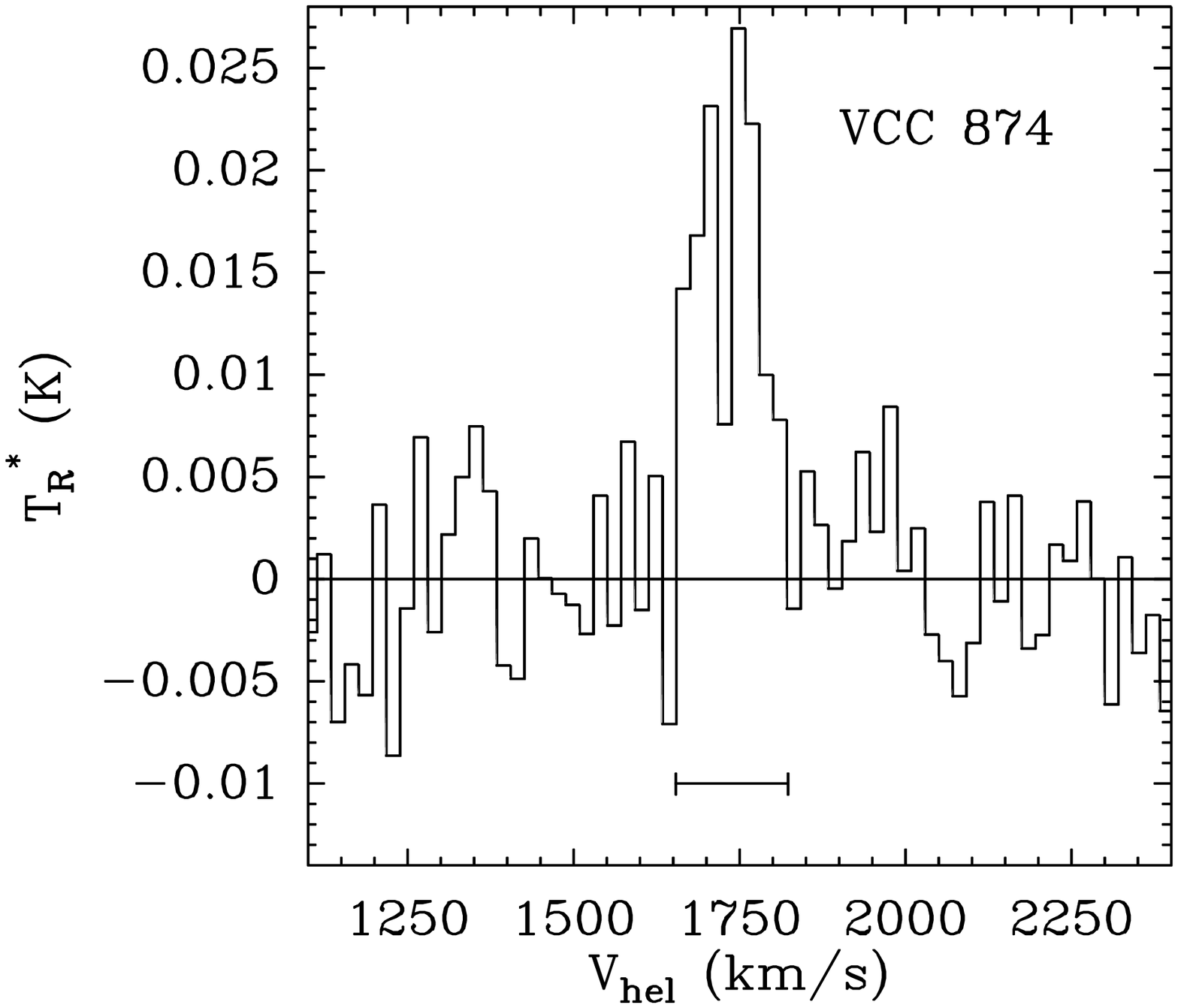}
\includegraphics[width=5.5cm,angle=-90,viewport=1 120 500 700,clip]{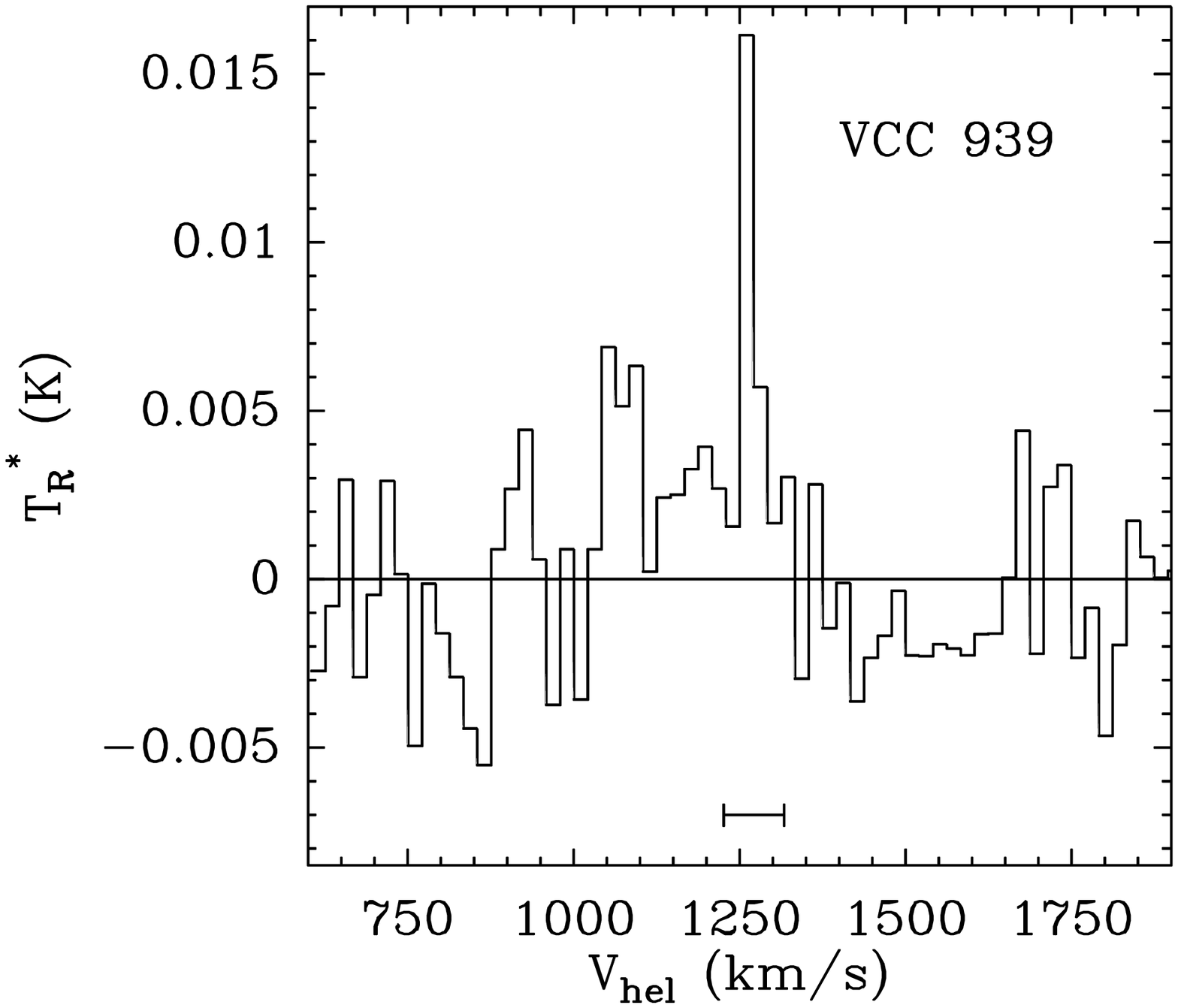}
\includegraphics[width=5.5cm,angle=-90,viewport=1 120 500 700,clip]{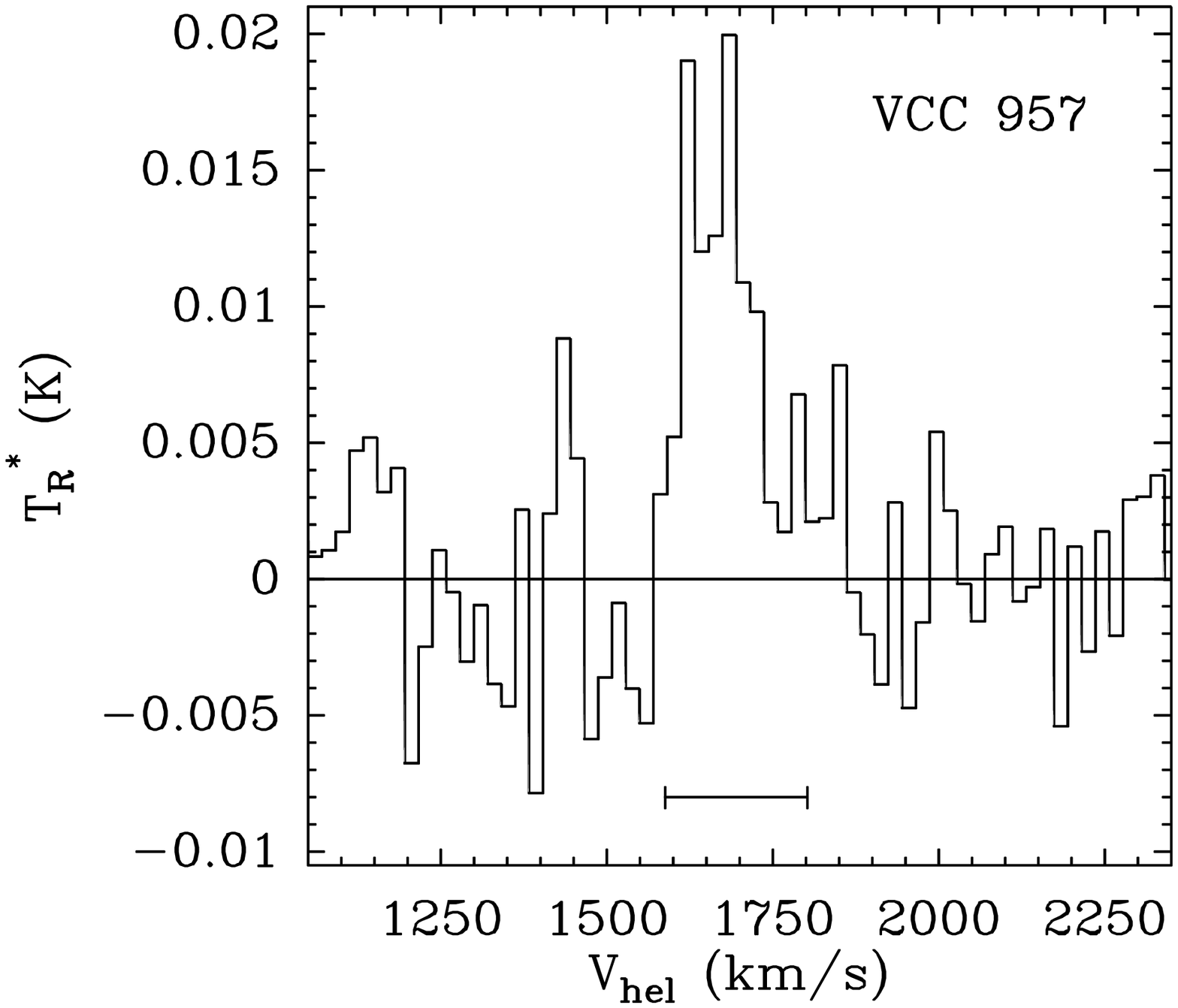}}
\centerline{
\includegraphics[width=5.5cm,angle=-90,viewport=1 120 500 700,clip]{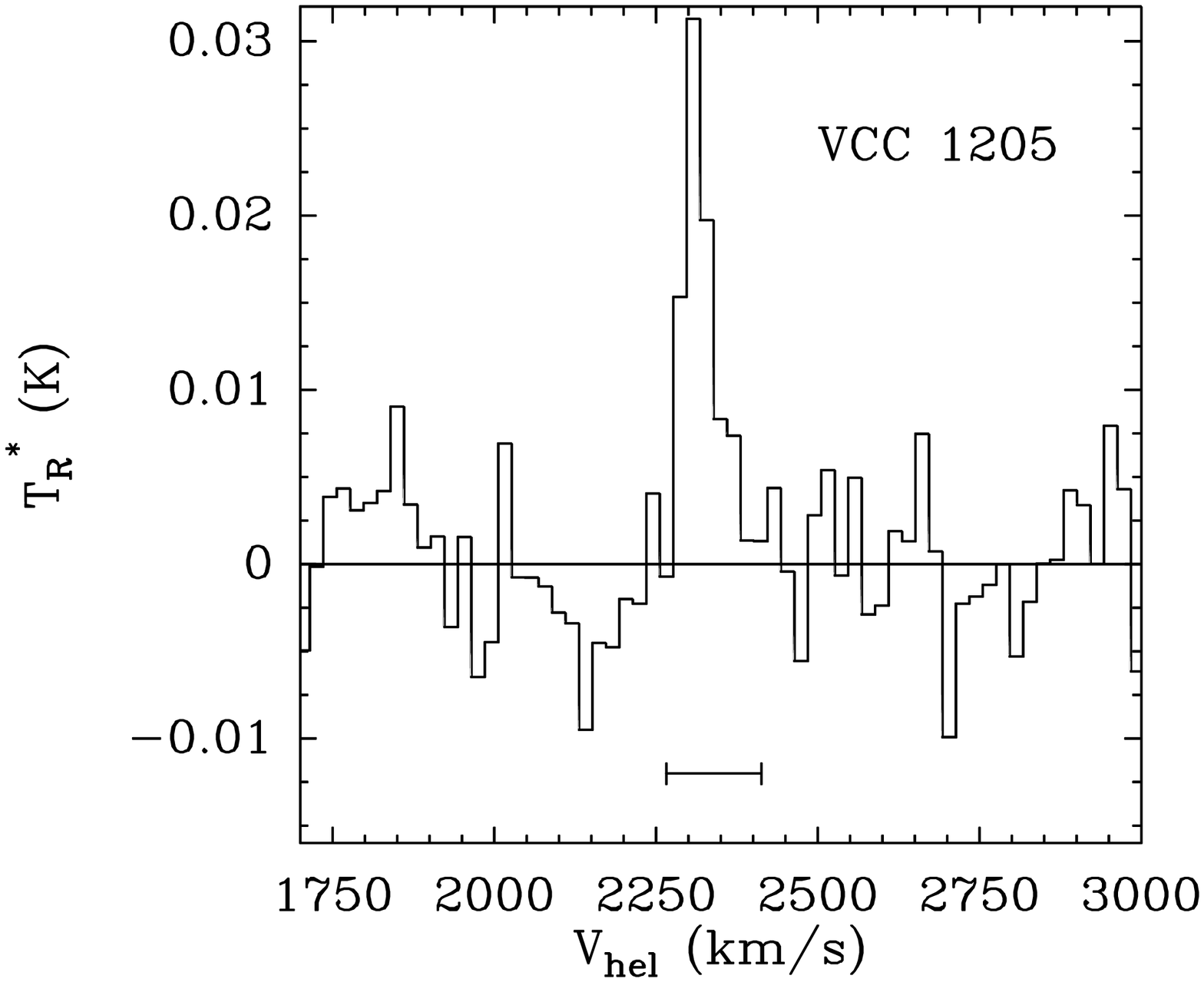}
\includegraphics[width=5.5cm,angle=-90,viewport=1 120 500 700,clip]{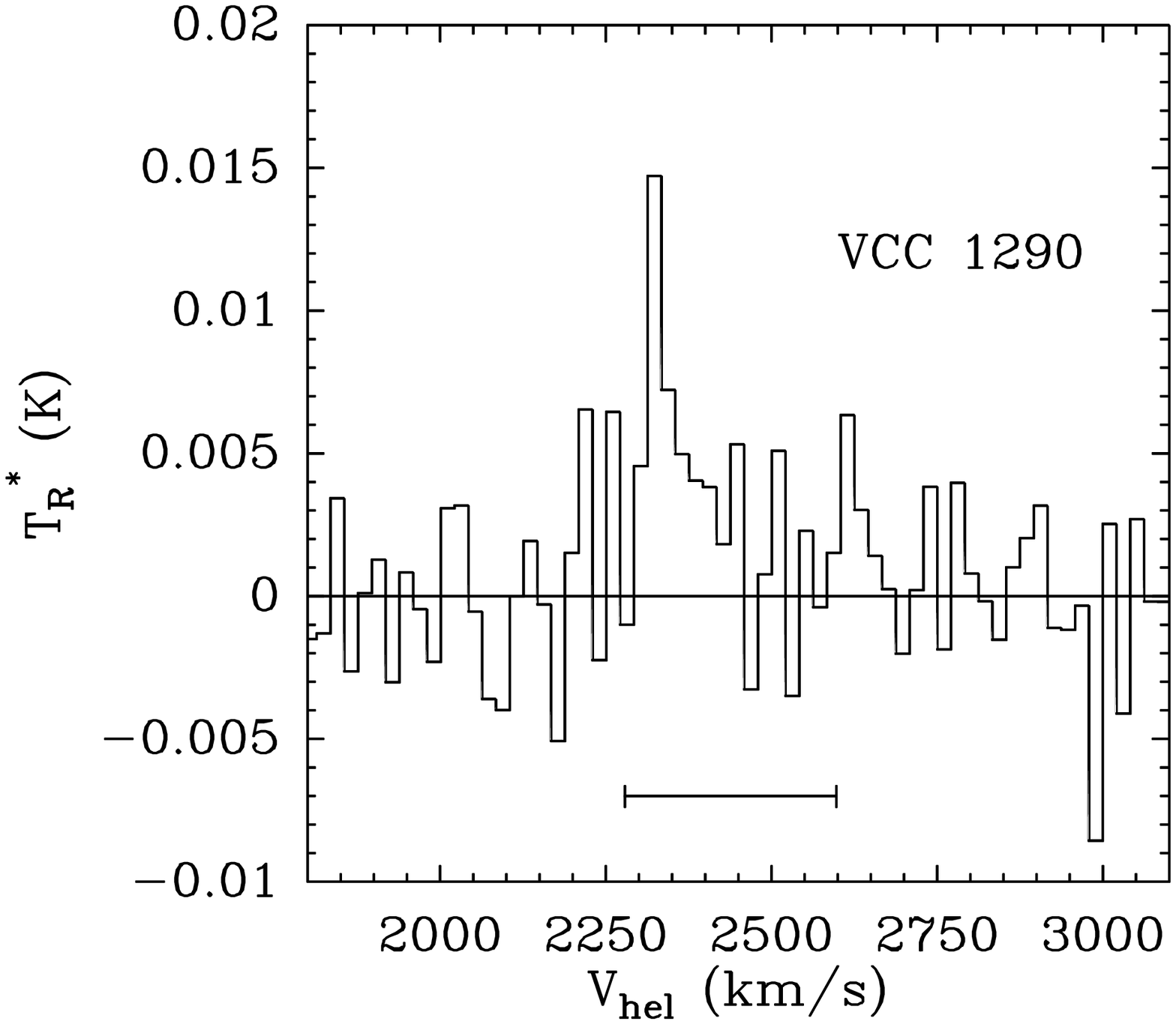}
\includegraphics[width=5.5cm,angle=-90,viewport=1 120 500 700,clip]{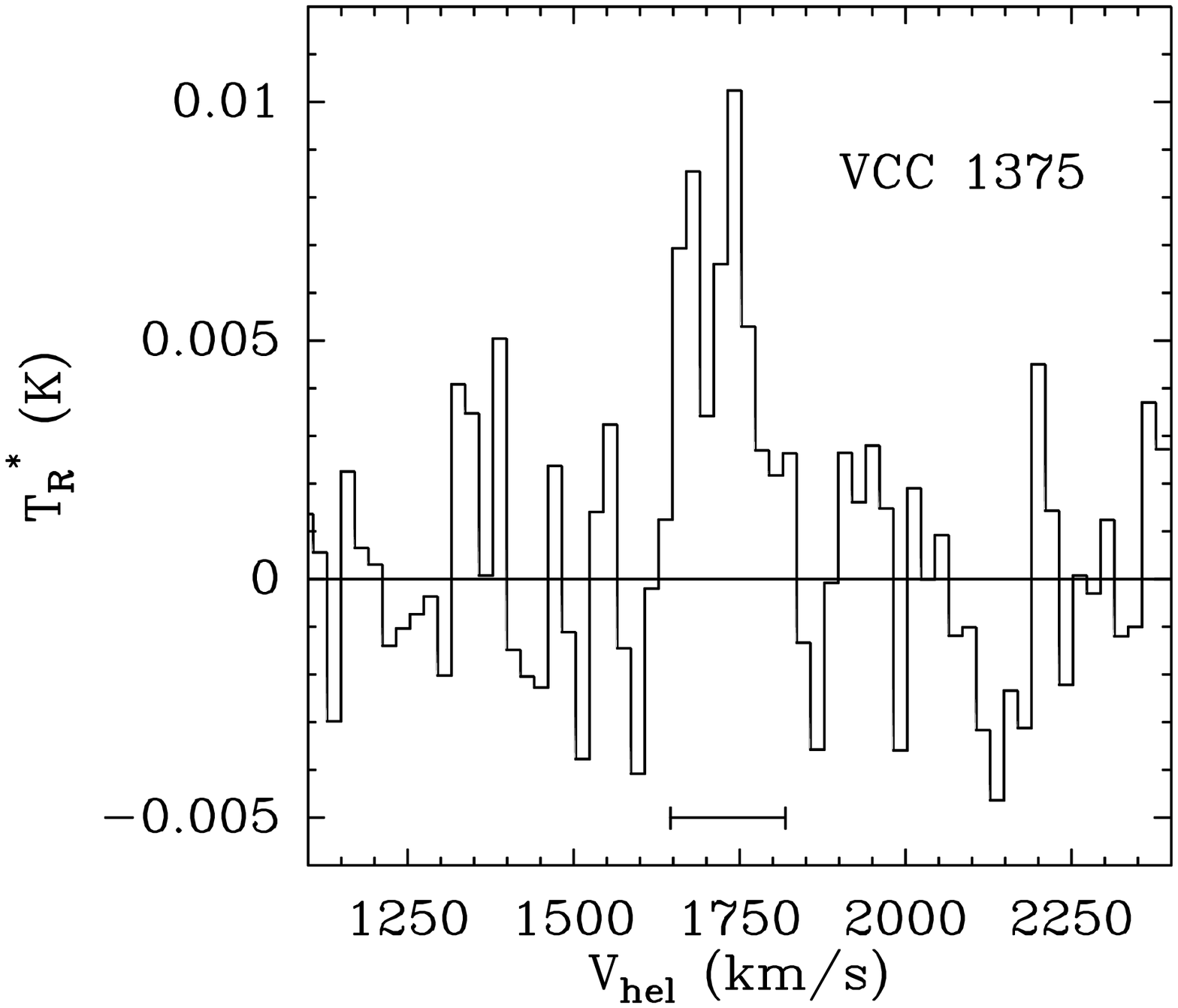}}
\centerline{
\includegraphics[width=5.5cm,angle=-90,viewport=1 120 500 700,clip]{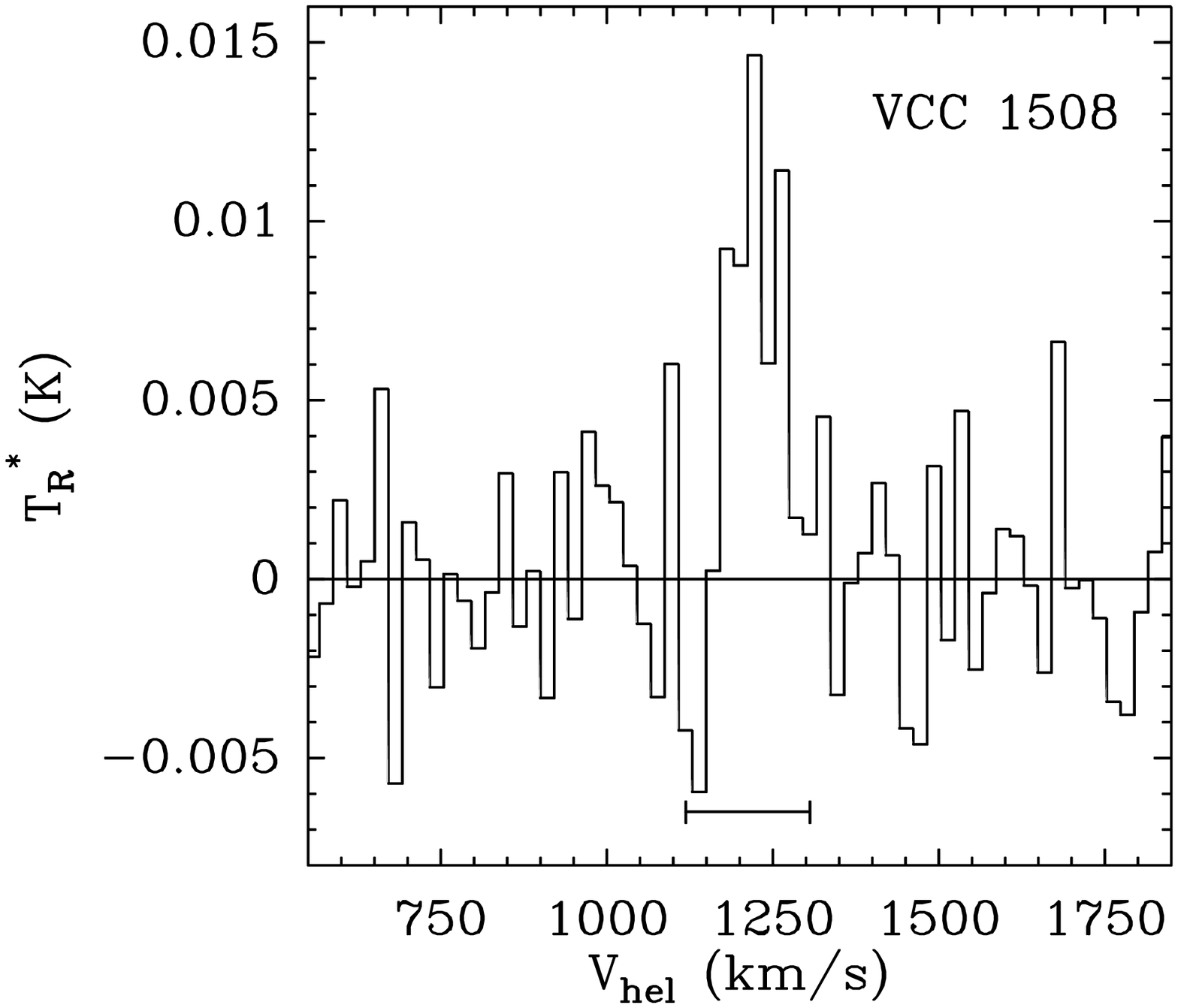}
\includegraphics[width=5.5cm,angle=-90,viewport=1 120 500 700,clip]{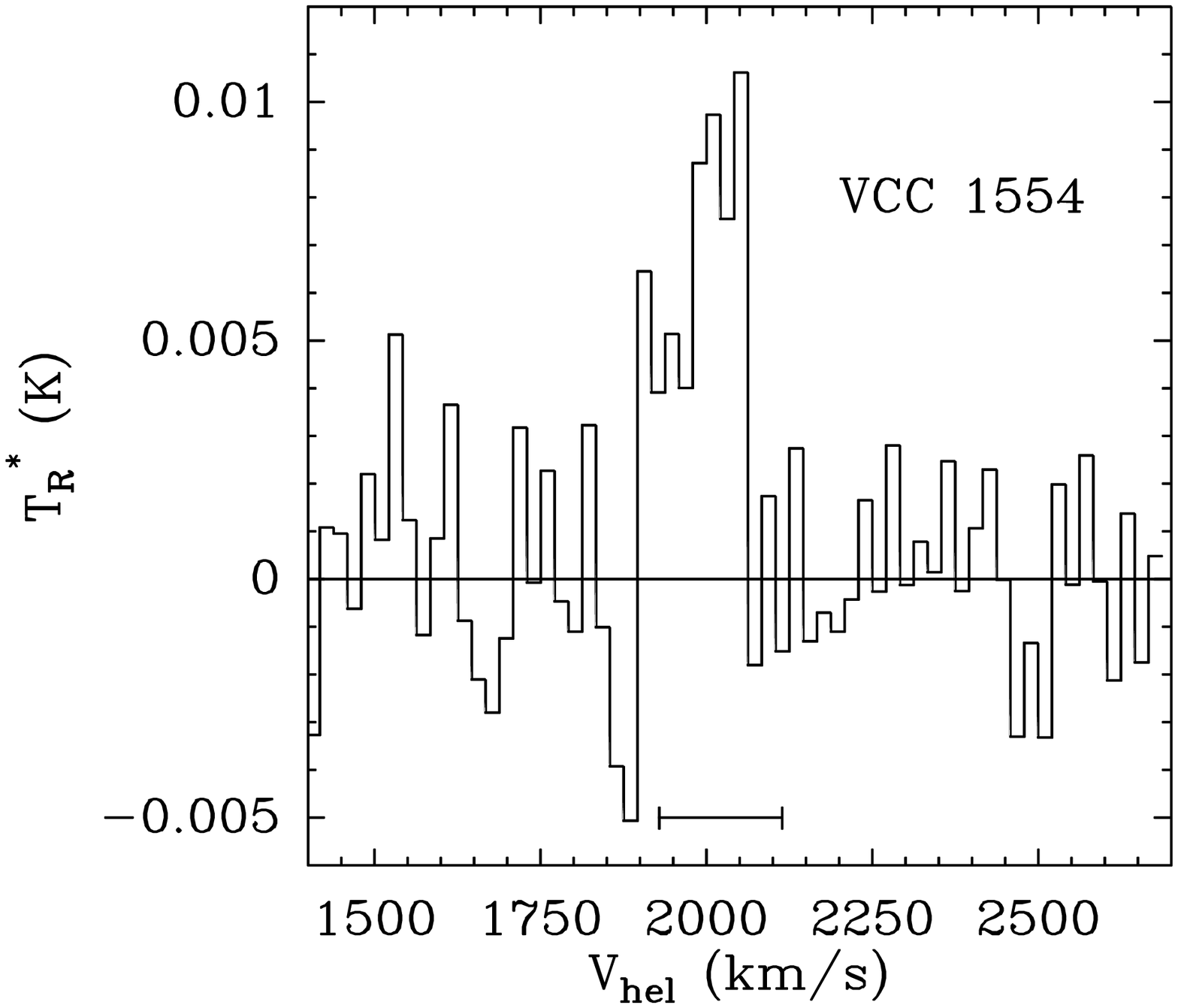}
}
\caption{$^{12}$CO(1-0) line spectra (smoothed to $\delta V_{CO}$=21 
km s$^{-1}$) of the 11 detected galaxies (in order of increasing VCC name) 
in the $T^*_R$ scale. The horizontal line indicates the HI line width 
listed in Table 1.}
\end{figure*}
\clearpage

\begin{equation}
{\Delta I(CO)=2 \sigma (\Delta V_{CO} \delta V_{CO})^{1/2} 
{\rm {K~km~s ^{-1}}}}
\end{equation}

where $\sigma$ is the rms noise of the spectrum, $\Delta V_{CO}$ 
is the CO linewidth 
(given in Col. 7), and $\delta V_{CO}$ is the spectral 
resolution. Spectra were 
smoothed to $\delta V_{CO}$ = 21 km s$^{-1}$.}
\item {Column 6: Heliocentric velocity determined from the CO line (gaussian 
fit), in km s$^{-1}$ (optical definition $v$=$cz$=$\Delta\lambda/\lambda_0$). 
The estimated error is 
comparable to the resolution, thus $\sim$ 20 km s$^{-1}$.}
\item {Column 7: Width at 50\% of the maximum intensity of the CO line, 
in km s$^{-1}$, with an estimated 
error of $\sim$ 20 km s$^{-1}$.}
\item {Column 8: The  $M(H_2)_{CO}$ mass, 
determined as in sect. 4.2 adopting a standard $X$ conversion factor
$X$= 2.3 10$^{20}$ mol cm$^{-2}$ (K km s$^{-1}$)$^{-1}$.}
\item {Column 9: Filling factor, defined as the ratio of the beam area to the 
optical surface of the galaxy.}
\item{Column 10: The molecular gas mass determined from the CO line intensity
as described in sect. 7
assuming the H luminosity-dependent $X$ conversion factor given in Table 5}  
\end{itemize}

\begin{table*}
\caption{Results of the observations, in the $T^*_R$ scale.}
\label{Tab2}
\[
\begin{array}{rrrrccrrlr}
\hline
\noalign{\smallskip}
name & int-time &  rms &  I(CO) &  \Delta I(CO) & V(CO) & \Delta V_{CO} & log M(H_2)_{CO} & FF & log M(H_2)\\
 & min(on+off) &\rm{mK}&\rm{K~km~s^{-1}} &\rm{K~km~s^{-1}} &\rm{km~s^{-1}} &\rm{km~s^{-1}} &\rm{M\odot} &  &\rm{M\odot}\\
\noalign{\smallskip}
\hline
\noalign{\smallskip}
  58 & 162 & 3 & <0.44 & -  & -   & -   & <8.04 & 0.19&<8.10\\
  87 & 156 & 4 & <0.38 & -  & -   & -   & <7.43 & 0.80&<8.15\\
  92 &  30 & 9 &  7.66 &1.56&-102 & 360 &  8.74 & 0.04& 8.49\\
  97 &  96 & 3 &  3.89 &0.55&2442 & 400 &  8.99 & 0.44& 8.91\\
 199 & 240 & 3 & <0.86 & -  & -   & -   & <8.21 & 0.29&<8.01\\
 318 & 132 & 3 & <0.37 & -  & -   & -   & <7.96 & 0.49&<8.37\\
 459 & 204 & 2 & <0.21 & -  & -   & -   & <7.17 & 2.78&<7.75\\
 792 & 192 & 2 &  1.86 &0.27& 934 & 210 &  8.38 & 0.14& 8.25\\
 874 & 120 & 4 &  2.53 &0.40&1736 & 120 &  8.25 & 0.40& 8.37\\
 939 & 240 & 3 &  0.52 &0.13&1265 &  20 &  7.83 & 0.07& 8.00\\
 957 & 174 & 4 &  2.30 &0.37&1669 & 100 &  8.21 & 0.49& 8.36\\
1205 & 120 & 4 &  1.72 &0.23&2312 &  40 &  8.09 & 0.40& 8.30\\
1290 & 198 & 3 &  1.23 &0.17&2336*&  40 &  7.94 & 0.39& 8.09\\
1375 & 270 & 2 &  0.93 &0.21&1720 & 130 &  7.81 & 0.05& 8.17\\
1412 & 114 & 7 & <1.07 & -  & -   & -   & <7.88 & 0.11&<7.81\\
1508 & 408 & 3 &  1.08 &0.29&1226 & 110 &  7.88 & 0.09& 8.02\\
1554 & 222 & 2 &  1.17 &0.24&2006 & 170 &  7.92 & 0.32& 8.08\\
1686 & 168 & 4 & <0.39 & -  & -   & -   & <7.45 & 0.18&<7.81\\
1929 & 228 & 3 & <0.38 & -  & -   & -   & <7.43 & 0.32&<7.72\\
1943 &  78 & 10& <1.57 & -  & -   & -   & <8.05 & 0.13&<7.70\\
2023 & 282 & 4 & <0.50 & -  & -   & -   & <7.55 & 0.42&<7.96\\
N4866& 168 & 4 & <0.85 & -  & -   & -   & <7.78 & 0.10&<7.67\\ 
\noalign{\smallskip}
\hline
\end{array}
\]
*: poor quality fit
\end{table*}

\subsection {Comparison with previous observations}

CO measurements are often used to estimate the molecular hydrogen 
content of galaxies adopting a standard conversion factor from CO 
intensities to H$_2$ column densities ($X$). In order to allow a comparison with 
previous works we convert the CO line fluxes into H$_2$ masses 
using the conventional Galactic conversion factor 
$X$=2.3 10$^{20}$ mol cm$^{-2}$ (K km s$^{-1}$)$^{-1}$ 
of Strong et al. (1988) where $I(CO)$ from Column 4 of Table 
2 is converted into the $T_{mb}$ scale (see sect. 3). 
The molecular gas mass ($M(H_2)_{CO}$, in solar units) as determined from CO measurements, is given by:

\begin{equation}
{M(H_2)_{CO}={\rm 68.08 \times} ({D/Mpc})^2 I(CO) (K km s^{-1}) (\theta/1")^2}
\end {equation}

\noindent
where $\theta$ is the half-power beam width (HPBW) of the telescope and $D$ the 
distance to the source (as given in Table 1). 

\noindent
Since the galaxies were 
observed only at the central position, our mass determinations should be 
considered as lower limits of the total $M(H_2)_{CO}$. However since the optical angular 
sizes of the observed galaxies do not generally exceed the adopted beam 
by more than a factor of two, and since the CO emission in spiral 
galaxies is centrally peaked, with an exponential distribution
with a scale length $\approx$ 1.5 smaller than the optical one 
(Young et al. 1995), the present data should 
give reliable estimates of the total CO emission. \\
Seven target galaxies were previously observed using the 13.7m FCRAO radiotelescope,
which, at this frequency, has a beam size of 45 arcsec, thus comparable 
to the NRAO 12m. 
We compare in Table 3 the different 
sets of data. The intensities given in Table 3 are transformed to the main beam scale
adopting a main beam efficiency $\eta$$_{mb}$=0.53 for the FCRAO.
Masses are estimated adopting the same CO to H$_2$ standard conversion factor 
$X$=2.3 10$^{20}$ mol cm$^{-2}$ (K km s$^{-1}$)$^{-1}$
and distance
(as given in Table 1). For galaxies observed at several positions, we compare 
the central beam observations.

\begin{table*}
\caption{Comparison between our data ($TW$) and those available in the literature ($L$).}
\label{Tab3}
\[
\begin{array}{rcrcrrc}
\hline
\noalign{\smallskip}
name & rms &  I(CO)      & \delta V_{CO} & log M(H_2)_{CO_{TW}} & log M(H_2)_{CO_L} & reference \\
     &\rm{mK}&\rm{K~km~s^{-1}}&\rm{km~s^{-1}}&\rm{M\odot}       &\rm{M\odot}        &           \\
\noalign{\smallskip}
\hline
\noalign{\smallskip}
  92 & 13  &  9.29       & 12            &  8.74             &  8.57        & KY        \\
 792 & 15  & <1.65       & 12            &  8.38             & <8.08        & KY        \\
1412 & 20  & <2.63       & 15            & <7.88             & <8.02        & Y         \\
1508 & 20  & <1.90       & 12            &  7.88             & <7.88        & K         \\
1554 & 14  & <1.32       & 12            &  7.92             & <7.72        & KY        \\
1943 & 10  & <1.20       & 12            & <8.05             & <7.68        & KY        \\
N4866& 12  & <1.95       & 12            & <7.78             & <7.89        & KY        \\
\noalign{\smallskip}
\hline
\end{array}
\]
KY: Kenney \& Young (1988a);
Y: Young et al. (1995); 
K: Kenney, private communication
\end{table*}

In spite of the different beam sizes of the telescopes, the two sets of CO intensity 
determinations are consistent within a factor of 2.
Our observations are generally more sensitive than the published ones, except for 
VCC 92 (NGC 4192), which was observed as a cross check, and VCC 1943 (NGC 4639), which was
interrupted because of incoming bad weather.

\section {Data analysis}
 
\subsection {The samples}

The molecular gas properties of late-type galaxies are analysed in this section using the presently reported
 data jointly with those available in the literature for a large sample of optically selected
galaxies in the Coma/A1367 supercluster and in the Virgo cluster (Boselli et al. 1997b;
1995a and references therein) extracted respectively from the Zwicky catalogue (CGCG)
and from the Virgo cluster catalogue of Binggeli et al. (1985, VCC). 
The sample, which is not complete in any sense, 
includes 266 normal galaxies with $^{12}$CO(1-0) data spanning a large range in morphological
type (Sa to Im and BCD) and luminosity (-16 $>$ $M_B$ $>$ -22).\\ 
The accuracy of the morphological classification is excellent for the 
Virgo galaxies (Binggeli et al. 1985; 1993).
Because of the higher distance, the morphology of galaxies belonging 
to the other surveyed regions suffers from an uncertainty 
of about $\pm$0.75 Hubble type bins.\\
We assume a distance of 17 Mpc for the members (and possible members) 
of Virgo cluster A, 22 Mpc for 
Virgo cluster B, 32 Mpc for
objects in the M and W clouds (see Gavazzi et al. 1999b).
Members of the Coma and A1367 clusters are assumed to be at 
distances of 86.6 and 92 Mpc respectively.  
Isolated galaxies in the Coma supercluster are assumed to be 
at their redshift distance adopting $H_o$ = 75 km s$^{-1}$ Mpc$^{-1}$.\\
Three different subsamples can be extracted from the Coma/A1367 supercluster 
and Virgo cluster galaxies:\\
i) the $unperturbed$ $sample$ is composed of 153 late-type galaxies, both isolated or
cluster members, whose HI-deficiency 
(defined as the logarithm of the ratio of the HI mass to the 
average HI mass of isolated objects
of similar morphological type and linear size (Haynes \& Giovanelli 1984)) is $HI-def$ $\leq$ 0.3.\\
ii) the $isolated$ $sample$ is composed of 47 strictly isolated galaxies  in
the bridge between Coma and A1367 (see Gavazzi et al., 1999a).\\
iii) the $ISOPHOT$ $sample$ consists of 18 galaxies in Coma and A1367 observed by ISOPHOT (see next section;
Contursi et al. 2001).\\
A small sample of 14 nearby galaxies
with an independent measure of the $X$ conversion factor is described in Table 4.

\subsection {The complementary data}

Multifrequency spectrophotometric data are available for most of the analysed galaxies.\\
The CO data (266 sources) are taken from
Boselli et al. (1995a, 1997b), Kenney \& Young (1988a), Young et al. (1995) 
and references therein. For mapped galaxies (most of the bright Virgo spirals),
the total CO emission is estimated as the sum of the emission in each single
pointing. The error on the CO line intensity is approximately $\pm$ 10 \%.\\
211 of the sample galaxies have been observed in H$\alpha$.
H$\alpha$+[NII] fluxes obtained from imaging, 
aperture photometry or integrated spectra 
are taken from Kennicutt \& Kent (1983), Kennicutt et al. (1984), 
Gavazzi et al. (1991) Gavazzi et al. (1998), Moss et al. (1998),  
and references therein. Additional observations of several galaxies have been 
recently obtained by us 
during several runs at the Observatoire d'Haute Provence (France), at San Pedro
Martir (Mexico) and at Calar Alto (Boselli \& Gavazzi 2002; Gavazzi et 
al. 2002).
$H\alpha+[NII] E.W.$ (equivalent widths) from Kennicutt \& Kent (1983), 
have been multiplied by 
1.16, as suggested by Kennicutt et al. (1994), in order to account
for the continuum flux overestimate due to inclusion of the telluric 
absorption band near 6900 \AA ~in the comparison filter.
The estimated error on the $H\alpha+[NII] E.W.$ is $\sim$ $\pm$ 7\%.\\
HI fluxes are available for 263 galaxies.
HI data are taken from Scodeggio \& Gavazzi (1993) and  
Hoffman et al. (1996) and references therein.
HI fluxes are transformed into neutral hydrogen masses
with an uncertainty of $\sim$ $\pm$ 5\%.\\
NIR data for 255 galaxies, mostly from Nicmos3 observations, are taken from  
Gavazzi et al. (1996a,b, 2000a), Boselli et al. (1997a, 2000).
From these data we derive total (extrapolated to infinity)
magnitudes $H_T$, as described in Gavazzi et al. (2000b)
with typical uncertainties of $\sim$ $\pm$ 5 \%. These are 
converted into total luminosities 
using: $log L_H = 11.36 - 0.4H_T +2logD$ 
(in solar units), 
where $D$ is the distance to the source (in Mpc). For a few objects we
derive the H luminosity
from K' band measurements assuming an average H-K' colour of
0.25 mag (independent of type; see Gavazzi et al. 2000a).
A minority of the objects in our sample have an H band magnitude
obtained from aperture photometry, thus with no asymptotic extrapolation.
For these we use the magnitude $H_{25}$ determined as in Gavazzi \& Boselli 
(1996) at the optical radius
(the radius at which the B surface brightness is 25 mag arcsec$^{-2}$)
which is on average 0.1 magnitudes fainter than $H_T$ (Gavazzi et al. 2000a,b).
The total H magnitudes are corrected 
for internal extinction according to Gavazzi \& Boselli (1996).
No such correction has been applied to galaxies of type $>$ Scd.\\
Metallicity measurements $12 +log(O/H)$ are available for a small fraction of the sample (46 galaxies).
These are either determined at $r$$=$ 0.4$\rho_0$ for galaxies 
from individual HII region measurements 
(Zaritsky et al. 1994), or average estimates
if obtained from integrated spectra (Kennicutt 1992, Gavazzi et al. in
preparation). \\
Far-IR fluxes at 100 $\mu$m from IRAS for 262 galaxies are taken from several compilations
such as Bicay \& Giovanelli (1987), Thuan \& Sauvage (1992) and references therein.
For 18 galaxies ISOPHOT data in the wavelength range 100 - 200 $\mu$m are also available 
(Contursi et al. 2001). \\
The corresponding multifrequency spectrophotometric data for the reference 
sample of nearby galaxies are listed in Table 4. 
Giant molecular complexes of angular dimensions larger than 100 pc might 
be in non-equilibrium. Furthermore the contribution of the atomic hydrogen
inside these complexes to their total mass can be important.
An accurate estimate of the molecular gas mass using the virial theorem can be
obtained only for the dense cores in virial equilibrium of giant molecular 
clouds, where the contribution of HI is negligible. These cores have
linear dimensions of $\sim$ 10 pc. We thus decided to include in Table 4
only those galaxies with interferometric CO observations where the high 
resolution ($\le$ 100 pc) allowed the determination of $X$ from CO line
widths and intensities measurements of the resolved cores of molecular clouds.

\begin{table*}
\caption[]{Multifrequency data for the sample of nearby galaxies.}
\label{Tab4}
\[
\begin{array}{llllrcccccccc}
\hline
\noalign{\smallskip}
name & type & dist     &  M_B^a& est.~log L_H^b	& 12+log(O/H) 	& ref	& H\alpha+[NII] E.W. & ref & X 	& method^e &ref 	\\
     &      &\rm{Mpc}  &\rm{mag}&\rm{solar~units}&             	&     	&   \AA    	& & 10^{20} 	&	& \\
\noalign{\smallskip}
\hline
\noalign{\smallskip}
MW   &          &0	&-20.40	&10.76^c	&8.90 \pm 0.04	&1	&-		&-	&1.56 \pm 0.05	&gamma  &15	\\
SMC  &SB(s)m	&0.061	&-16.30	& 8.81^c	&8.04 \pm 0.06	&1	&24 \pm 5	&5	&10.0		&virial &16	\\
LMC  &SB(s)m	&0.055	&-18.40	& 9.81^c	&8.37 \pm 0.12	&1	&36 \pm 8	&5	&8.0		&virial &17	\\
M31  &SA(s)b    &0.77	&-20.67	&11.07		&9.01 \pm 0.10	&1	&3.7 \pm 1.4^d	&6	&2.4 \pm 1.5	&virial &1	\\
M33  &SA(s)cd	&0.84	&-18.31	& 9.90		&8.78 \pm 0.05	&2	&20 \pm 2	&7	&5.0 \pm 1.5	&virial &1	\\
M51  &SA(s)bc	&9.6	&-20.75	&11.08		&9.23 \pm 0.12	&2	&19 \pm 2.5	&8	&0.6		&mm     &16	\\
M81  &SA(s)ab   &3.8    &-20.51 &10.90          &9.00 \pm 0.13  &2      &10 \pm 4.5     &8	&0.7^f		&virial &18,19	\\
M82  &I0sbrst	&3.63	&-18.94	&10.69		&9.00 \pm 0.12	&3	&42 \pm 1	&8	&1		&mm     &20	\\
IC10 &IBm       &0.66	&-15.90	& 8.62^c	&8.31 \pm 0.20	&1	&-		&-	&6.6 \pm 2.2	&virial &1	\\
N891 &SA(s)b?Sp	&9.5	&-20.52	&10.78		&-		&-	&7.5 \pm 1	&10,11,12,13&1.5	&mm     &16,22	\\
N1569&IBm	&2.2	&-16.90	& 8.79		&8.19 \pm 0.02	&4	&149 \pm 15	&9,14	&15^f		&virial &10	\\
N4565&SA(s)b?Sp	&9.4	&-21.77	&11.13		&-		&-	&2.7 \pm 1.5^d	&12	&1.0		&mm     &16	\\
N6822&IB(s)m	&0.5	&-15.10	& 8.15		&8.16 \pm 0.06	&1	&-		&-	&6.6 \pm 3.9	&virial &1	\\
N6946&SAB(rs)cd &5.5    &-20.92 &10.60          &9.06 \pm 0.17  &2      &29 \pm 5       &14     &1.8            &mm     &21     \\
\noalign{\smallskip}
\hline
\end{array}
\]
a: B total corrected magnitudes are from NED, and distances from van den Bergh (1999); for the MW, $M_B$ is 
from van den Bergh (1999).

b: H band magnitudes, corrected for extinction as in Gavazzi \& Boselli (1996), are extracted from
the 2MASS survey and/or from aperture photometry (Gezari et al. 1993), or form recent observations taken at the 3.5 TNG
telescope (Gavazzi et al., in preparation). The estimated error on the H luminosity is $<$ $\pm$ 10 \% (photometric accuracy plus the
extrapolation to a total value), $\pm$ 0.12 mag (in log) when determined from the $L_H$ vs. $M_B$ relation given in c), 
and $\pm$ 0.15 mag for the Milky Way (MW). 
 
c: for galaxies not observed in the H band (5), the H luminosity
is determined from $log L_H$ (L$_H$ $\odot$)= 1.05 ($\pm$0.22) - 0.48($\pm$0.01)$\times$ $M_B$ (R$^2$=0.84, where R is the regression coefficient) 
has been determined from 246 galaxies in our sample. 

d: for M31 and NGC 4565 $H\alpha+[NII] E.W.$ have been estimated using published H$\alpha$ fluxes
and red continuum as determined from the R band photometry assuming 0.0 mag in R equivalent to 1.74 10$^{-9}$
erg cm$^{-2}$ s$^{-1}$. Given its high inclination, we corrected the H$\alpha$ flux of NGC 4565 for 2 magnitudes.   

e: $gamma$: from high-energy gamma-ray emission from EGRET observations; 
$virial$: from CO observations of resolved molecular clouds assuming virial 
equilibrium; $mm$: from millimetric observations assuming a 
metallicity-dependent dust to gas ratio.

f: uncertain values

References:

1: Arimoto et al. (1996), average of the disc;
2: Zaritsky et al. (1994);
3: Gavazzi et al., in preparation;
4: Kobulnicky \& Skillman (1997);
5: Kennicutt et al. (1995);
6: H$\alpha$ flux from Devereux et al. (1994), and R magnitude from Walterbos \& Kennicutt (1987);
7: Devereux et al. (1997), and private communication;
8: Boselli \& Gavazzi, in preparation;
9: Hunter et al. (1993);
10: Taylor et al. (1999);
11: Hoopes, private communication;
12: Rand et al. (1992);
13: Hoopes et al. (1999);
14: Kennicutt \& Kent (1983);
15: Hunter et al. (1997), average over the whole Milky Way;
16: Boselli et al. (1997b), scale of 10 pc for SMC;
17: Rubio 1999, scale 10 pc;
18: Taylor \& Wilson (1998);
19: Allen et al. (1997);
20: Wild et al. (1992), central 1 kpc;
21: Bianchi et al. (2000);
22: Alton et al. (2000)
\end{table*}

\section {An empirical calibration of $X$}

\subsection{The nearby sample of resolved galaxies}

Table 4 contains the best-estimated values of $X$ for a sample of 14 well-studied
nearby galaxies. 
Table 4 shows that $X$ spans the range
0.6 $\leq$ $X$ $\leq$ 10 10$^{20}$ mol cm$^{-2}$ (K km s$^{-1}$)$^{-1}$
for galaxies of different morphological type.\\ 
The ratio between $^{12}$CO(1-0) line emission and the molecular hydrogen
column density strongly depends on
several physical properties of the ISM such as the UV radiation field, 
the metallicity and the cosmic ray density (Lequeux et al. 1994; Kaufman et al. 1999), which are
known to vary from galaxy to galaxy. \\
We empirically quantify these dependences by plotting in Fig 2.
(left panels) the relationship between $X$, the
$H\alpha+[NII] E.W.$ and the metallicity ($12+log(O/H)$) for the 14
nearby galaxies.

\begin{figure*}
\centerline {
}
\vbox{\null\vskip 13. cm
\includegraphics{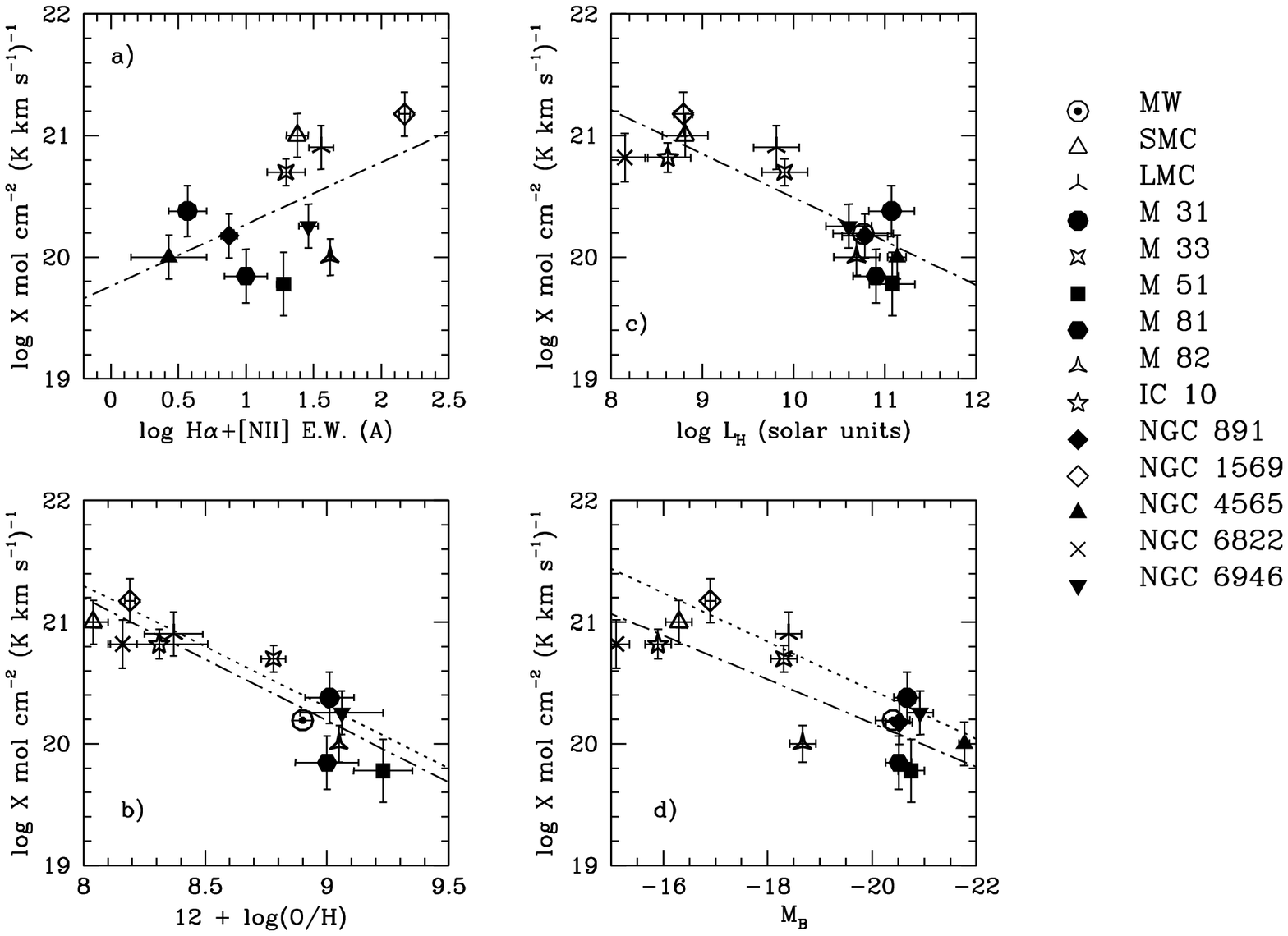}
}
\caption{The relationship for the template sample of nearby galaxies 
between the $X$ conversion factor from CO line intensity to H$_2$ column density and a) the $H\alpha+[NII] E.W.$, b) 
the metallicity index $12 + log(O/H)$, c) the H band luminosity and d) the absolute B magnitude. 
The dotted-dashed line is the
best fit to the data; the dotted line is the best fit given by Arimoto et al. (1996).
}
\label{fig.2}
\end{figure*}

\noindent
The relationship with the $H\alpha+[NII] E.W.$ (taken as a SFR tracer), 
if any, is ill-defined and that with metallicity is quite clear. 
In galaxies with a low metallicity
and a strong UV radiation field (high $H\alpha+[NII] E.W.$) (both go together in general) 
the $X$ conversion factor is 
a factor of $\sim$ 20 higher than in quiescent, high metallicity galaxies such as
the Milky Way. These two relationships can be used in principle to determine 
a more accurate value of the $X$ conversion factor once the metallicity and/or 
the $H\alpha+[NII] E.W.$ is known. Metallicity measurements 
are available for only a minority of galaxies, while $H\alpha+[NII] E.W.$ 
exist for a few hundred.
However there is a well-known anticorrelation (correlation) between  
$H\alpha+[NII] E.W.$ (metallicity) and luminosity (Gavazzi et al. 1998; Zaritsky et al. 1994)
in normal galaxies. It is reflected here by a strong relation between
$X$ and the H luminosity (Fig. 2c) or the B absolute magnitude (Fig. 2d).
The best fits to the data are given in Table 5.
The slope of the fits are consistent with those found by Arimoto et al. (1996), but significantely
steeper than that found by Wilson (1995) for the $X$ vs. $12 + log(O/H)$ relation
(see Table 5). This difference in slope with Wilson is probably due to the fact that our
sample includes many metal rich spiral galaxies with low values of $X$ 
($X$ $\leq$ 10$^{20}$ mol cm$^{-2}$ (K km s$^{-1}$)$^{-1}$)
not present in the Wilson's sample.
Our intercept for the $X$ vs. $12 + log(O/H)$ relation is consistent 
with that of Arimoto et al. (1996). 
Our intercept in the $X$ vs. $M_B$ relation is lower since
Arimoto et al. (1996) includes all the objects rejected here whose $X$
value is probably overestimated due to the low spatial resolution of the
CO observations ($>$ 100 pc).

\begin{table*}
\caption[]{
The relationships between the $X$ conversion factor and different galaxy properties.
$X$ is defined as:
log$X$ = slope$\times$Variable + constant (in units of mol cm$^{-2}$ (K km s$^{-1}$)$^{-1}$)
}
\begin{flushleft}
\begin{tabular}{lcccc}
\hline
This work &      &          &              &       \\
\hline
Variable & slope & constant & n.~of~objects& R$^2$$^a$ \\
\hline
$log H\alpha+[NII] E.W.^b$   &  0.51 $\pm$ 0.26 & 19.76 $\pm$ 0.43 & 11 & 0.29 \\ 
$12+log(O/H)$                & -1.01 $\pm$ 0.14 & 29.28 $\pm$ 0.20 & 12 & 0.83 \\
$log L_H^c$                  & -0.38 $\pm$ 0.06 & 24.23 $\pm$ 0.24 & 14 & 0.75 \\
$M_B$  		             &  0.18 $\pm$ 0.04 & 23.77 $\pm$ 0.28 & 14 & 0.67 \\
\hline
Literature &      &          &              &       \\
\hline
Variable & slope & constant & reference & \\
\hline
$12+log(O/H)$                & -1.00            & 29.30            & 1 & \\
$12+log(O/H)$                & -0.67 $\pm$ 0.10 & 26.43 $\pm$ 0.86 & 2 & \\
$M_B$                        &  0.20            & 24.44            & 1 & \\
\hline
\end{tabular} 
\end{flushleft}
a) regression coefficient

b) in logarithmic scale (in \AA)

c) in logarithmic scale (in solar units)

References:
1: Arimoto et al. (1996)
2: Wilson (1995)
\end{table*}

The relationships given in Table 5
between $X$ and $L_H$ and/or $M_B$ are de facto empirical calibrations
for a luminosity-dependent $X$ conversion factor. 
Given the large uncertainty in the determination of $X$ in the nearby
sample of galaxies, it is difficult to quantify the resulting accuracy in 
the molecular gas mass estimated using the relationships given in Table 5.  
We should also remind that even inside a given object $X$ might change by
a factor of $\sim$ 10 from the diffuse medium to the core of GMCs (Polk et al. 1988);
it is thus difficult to estimate a single value of $X$ representing the entire galaxy. 
We can however conclude that the adoption of the relations given in Table 5
should remove the first-order systematic effect with luminosity
in the estimate of the molecular hydrogen content of galaxies using CO data. 
The use of a standard $X$ conversion factor as those generally used in the literature 
($X$=2.3-2.8 10$^{20}$ mol cm$^{-2}$ (K km s$^{-1}$)$^{-1}$)
overestimates the molecular gas mass by a factor of $\sim$ 2-3
in massive galaxies of $L_H$ $\sim$ 10$^{11}$ L$_H{\odot}$, or $M_B$ $\sim$ -20.5 mag,
while underestimates M(H$_2$) in low mass objects of $L_H$ $\sim$ 10$^{9}$ L$_H{\odot}$, 
or $M_B$ $\sim$ -17 mag as those observed in this work by a factor of $\sim$ 2.
The relationship between $X$ and $12 + log(O/H)$ might
be used to estimate the radial distribution of molecular hydrogen in
galaxies mapped in CO with available measurements of the metallicity gradient.

\subsection{An alternative method}

An alternative technique for 
determining the molecular gas content can be pursued by 
assuming a metallicity-dependent 
dust to gas ratio and determining the
dust mass using far-IR or submillimetric continuum data.
The dust to gas ratio is then determined in regions with no CO emission, hence supposed 
to be strongly dominated by HI. In regions with CO emission, the excess dust emission 
with respect to this ratio indicates the mass of H$_2$.
This technique has been succesfully applied to M51, NGC 891, NGC 4565 and to some 
nearby irregular galaxies such as the Magellanic Clouds
(Gu\'elin et al. 1995; Gu\'elin et al. 1993; Neininger et al. 1996; Israel 1997).\\
In normal galaxies such as those in our sample the dust mass is dominated
by the cold dust emitting in the far-IR with a peak at $\sim$ 200 $\mu$m.
The determination of the total dust mass can be achieved provided that the 
100-1000 $\mu$m far-IR flux and the cold dust 
temperature are known. Recent observations aimed at determining the
spectral energy distribution in the far-IR of normal, quiescent
galaxies indicate that their SED can be fitted 
by a modified Planck law $\nu$$^{\beta}$$B_{\nu}(T_d)$, with $\beta$=2 (Alton et al. 2000).
The total dust mass can be thus determined from the relation (Devereux \& Young 1990):

\begin{equation}
{M_{dust}=CS_{\lambda}D^2(e^{a/T_{dust}}-1)    ~~~~{\rm M\odot}}
\end{equation}

\noindent
where $C$ is a quantity which depends on the grain opacity, $S_{\lambda}$
is the far-IR flux at a given wavelength (in Jy), $D$ is the distance of the 
galaxy (in Mpc), $T_{dust}$ is the dust temperature, and $a$ is a quantity which 
depends on $\lambda$. The majority of our sample has only IRAS data at 60 and 100 $\mu$m.
Given the strong contamination of the emission at 60 $\mu$m by very small grains, the 
60 to 100 $\mu$m ratio cannot be used to measure $T_{dust}$ (Contursi et al. 2001).
The ISOPHOT data at 100 and 200 $\mu$m give a better
measure of the dust temperature. $T_{dust}$ determined for the
ISOPHOT sample, as well as for 6 quiescent galaxies observed with ISOPHOT by Alton et
al. (1998) consistently with Contursi et al. (2001), seems to be independent of the 
UV radiation field as traced by the $H\alpha+[NII] E.W.$, of the metallicity or of 
the total luminosity. The average value is $T_{dust}$=20.8$\pm$3.2 K, which we
will assume also for galaxies without ISOPHOT measurements.
We then estimate the dust mass of the sample galaxies 
using eq. (4) with $C$=1.27 M$_{\odot}$Jy$^{-1}$Mpc$^{-2}$, consistent with Contursi et al. (2001), and 
$a$=144 K for $S_{\lambda}$=$S_{100 \mu m}$ 
(Devereux \& Young 1990).\\
The determination of the dust to gas ratio in a way consistent with that
obtained in the solar neighbourhood, requires the estimate of the gas and dust surface densities,
thus of the spatial distribution of dust and gas over the discs.
Unfortunately only integrated HI and dust masses are available for our spatially
unresolved galaxies. It is however reasonable to assume that the cold dust
is as extended as the optical disc (Alton et al. 1998).
The HI gas surface density is available only for a few galaxies in our sample
from VLA observations (Cayatte et al. 1994). For these objects we 
observe a good relationship between the HI surface density $\Sigma HI$ and 
the HI-deficiency parameter ($HI-def$), (defined as in sect. 5.1) (Fig. 3):

\begin{displaymath}
{log \Sigma HI =}
\end{displaymath}
\begin{equation} 
{= 20.92 (\pm 0.17)  - 0.65 (\pm 0.11) \times (HI-def)  ~~~~~~~~\rm{cm^{-2}}}
\end{equation}

\noindent
which can be used to predict $\Sigma HI$ for most of the galaxies of
our sample with single dish HI observations.

\begin{figure}
\vbox{\null\vskip 9. cm
\includegraphics{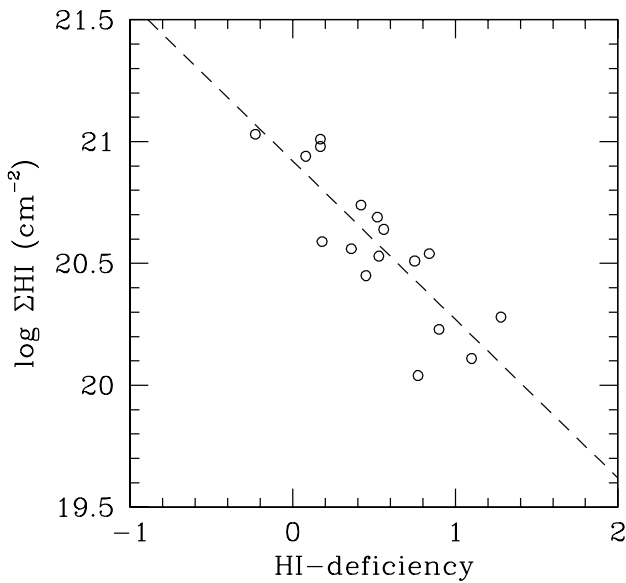}
}

\caption{The relationship between the HI surface density and the HI-deficiency
parameter for the galaxies in common with Cayatte et al. (1994). The dashed 
line gives the best fit to the data.
}
\label{fig.3}
\end{figure}

\noindent
The gas to dust ratio should depend on metallicity since 
the dust content is expected to be proportional to the metal content.
A gas to dust vs. metallicity relation can be calibrated using the 
data available for the MW (Sodroski et al. 1994), the LMC (4 times
solar; Koornneef 1982) and the SMC (10 times solar; Bouchet et al. 1985).
This gives the relation:

\begin{displaymath}
{log(gas/dust) = 10.207 (\pm 0.015) }
\end{displaymath}
\begin{equation} 
{- 1.146 (\pm 0.024) \times (12 + log(O/H)) + log(gas/dust)_{\odot}} 
\end{equation} 

\noindent
where the gas to dust ratio is given relative to the solar neighborhood,
estimated by Sodroski et al. (1994) at $(gas/dust)_{\odot}$=160.
The metallicity can be predicted using the metallicity vs. H band luminosity relation shown in Fig. 4:
\begin{equation} 
{12 + log(O/H) = 4.98 (\pm 0.15) + 0.37 (\pm 0.04) \times logL_H}
\end{equation} 

\noindent
The dispersion in the nearby sample is lower than in the other galaxies
probably because of a more accurate determination of the metallicity,
which has been measured from spectroscopic observations of single HII 
regions, while for the remaining galaxies it has been determined from 
long slit spectra integrated over their disc (drifting technique, 
Gavazzi et al. in preparation).
The adopted fit is however consistent with the result of Gavazzi et al., 
based on a large sample of galaxies with spectroscopic measurements.
\begin{figure}
\vbox{\null\vskip 9. cm
\includegraphics{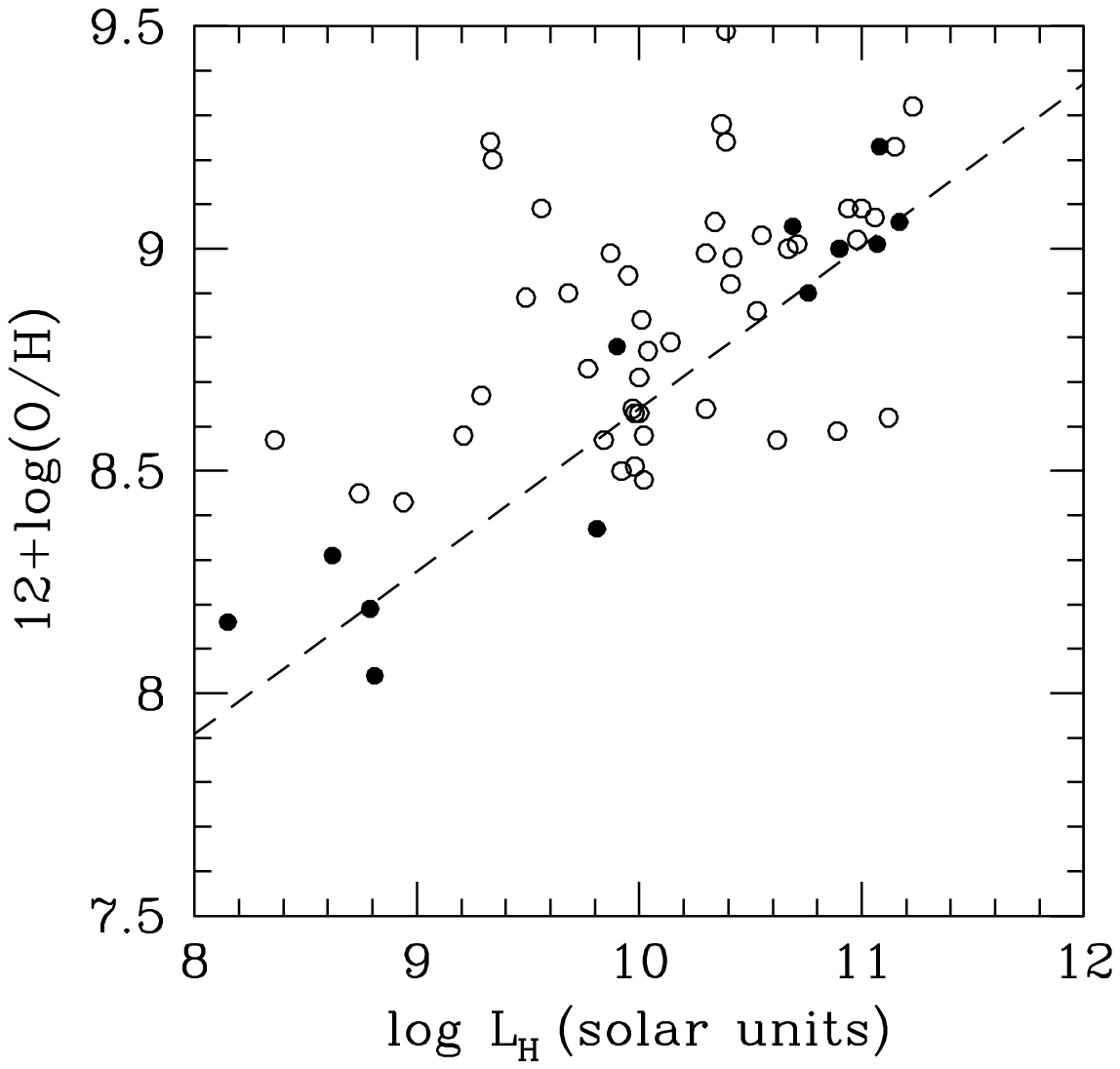}
}
\caption{The relationship between the metallicity ($12 + log(O/H)$) and
the H band luminosity. Filled dots are for the nearby galaxies, empty dots
for the remaining objects with metallicity measurements.  The dashed 
line gives the best fit to the data for the nearby sample.
}
\label{fig.4}
\end{figure}

\noindent
Using eq. (6) and (7) we can predict the gas to dust ratio for a galaxy
of any H luminosity.
The molecular gas mass comes directly if we assume that the H$_2$ is 
homogeneously distributed over the optical disc and
$X$ is given by $X$=$M(H_2)_{dust}$/$I(CO)$. The assumption of a homogeneous, 
flat distribution for the molecular hydrogen component over the disc of galaxies, 
which is in contradiction with the observational evidence that the CO emission 
is generally centrally peaked (see sect. 4.1), might introduce a systematic error in the
determination of $X$. We remark however that the expected H$_2$ distribution is
flatter than that of the CO emitting gas because of the observed decrease of the metallicity 
in the outer parts of galaxy discs.\\
\begin{figure*}
\vbox{\null\vskip 13. cm
\includegraphics{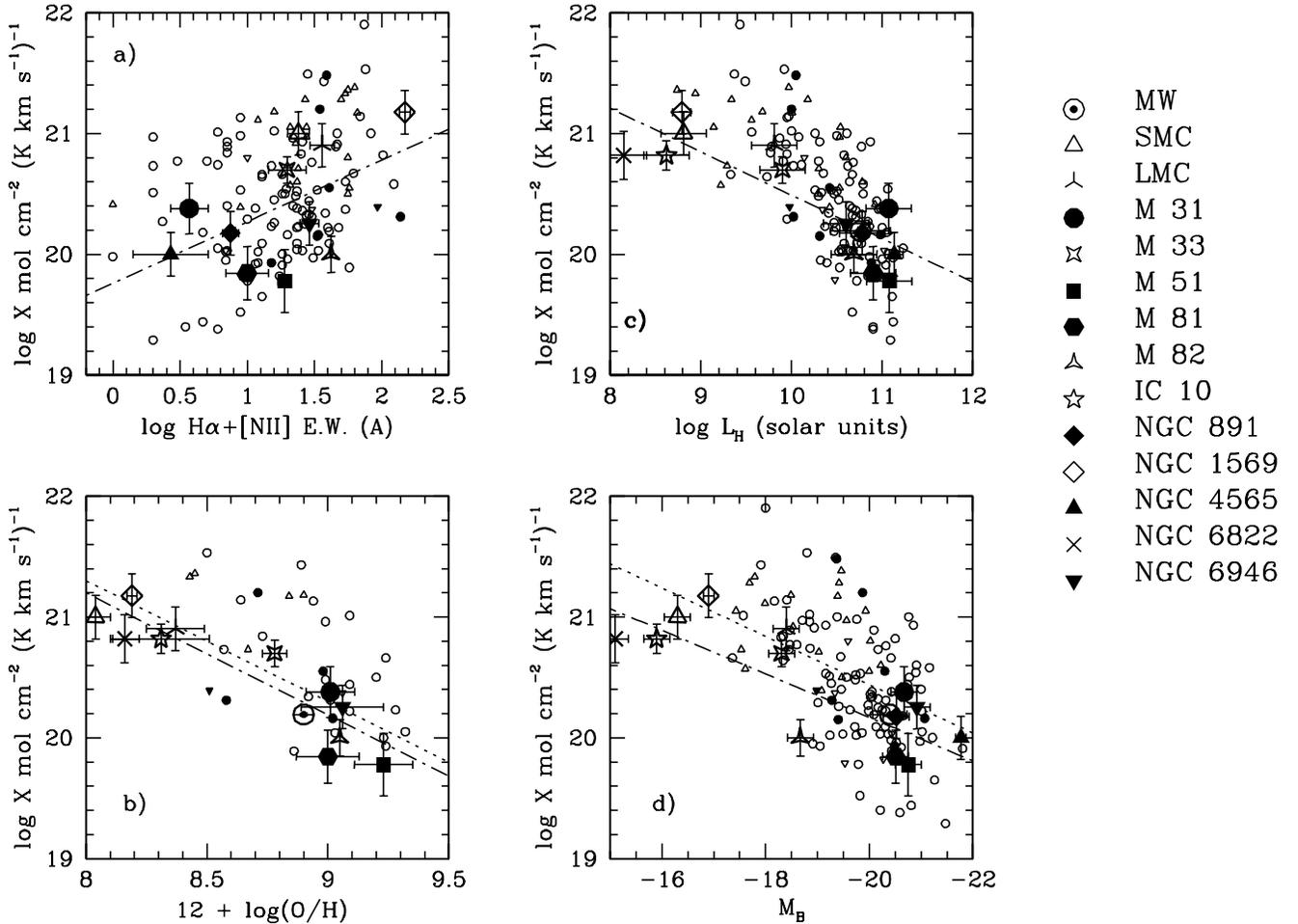}
}
\caption{Same as Fig. 2 but including values of $X$ determined using 
the alternative method described in Sect. 6.2 for the 
unperturbed sample (small symbols). 
Small open dots are for galaxies detected at 100 $\mu$m and CO, small open triangles
for galaxies undetected at 100 $\mu$m ($\bigtriangledown$) or CO ($\triangle$). Filled symbols are
for the ISOPHOT sample. 
}
\label{fig.5}
\end{figure*}
The values of $X$ obtained for the sample galaxies are compared with
those of the template galaxies in Fig. 5 (small symbols).
In spite of the larger scatter, it is encouraging to see that the new
values of $X$, at any given
luminosity, metallicity and UV radiation field, are in rough agreement with those
obtained for the template. The large uncertainty and systematic effects are not  
unexpected given the number of assumptions underlying the method. 
From Fig. 5 we conclude that the luminosity-dependent
$X$ conversion factor given in Table 5 is appropriate for estimating 
the molecular gas content of late-type galaxies
from $^{12}$CO(1-0) line intensity measurements. 
\noindent
From now on, the molecular gas content of the 266 sample galaxies,
$M(H_2)$, is estimated using the H band luminosity-dependent
$X$ value given in Table 5.

\section {Discussion}

\subsection{The molecular gas content of late-type galaxies} 

\begin{figure*}
\vbox{\null\vskip 12. cm
\includegraphics{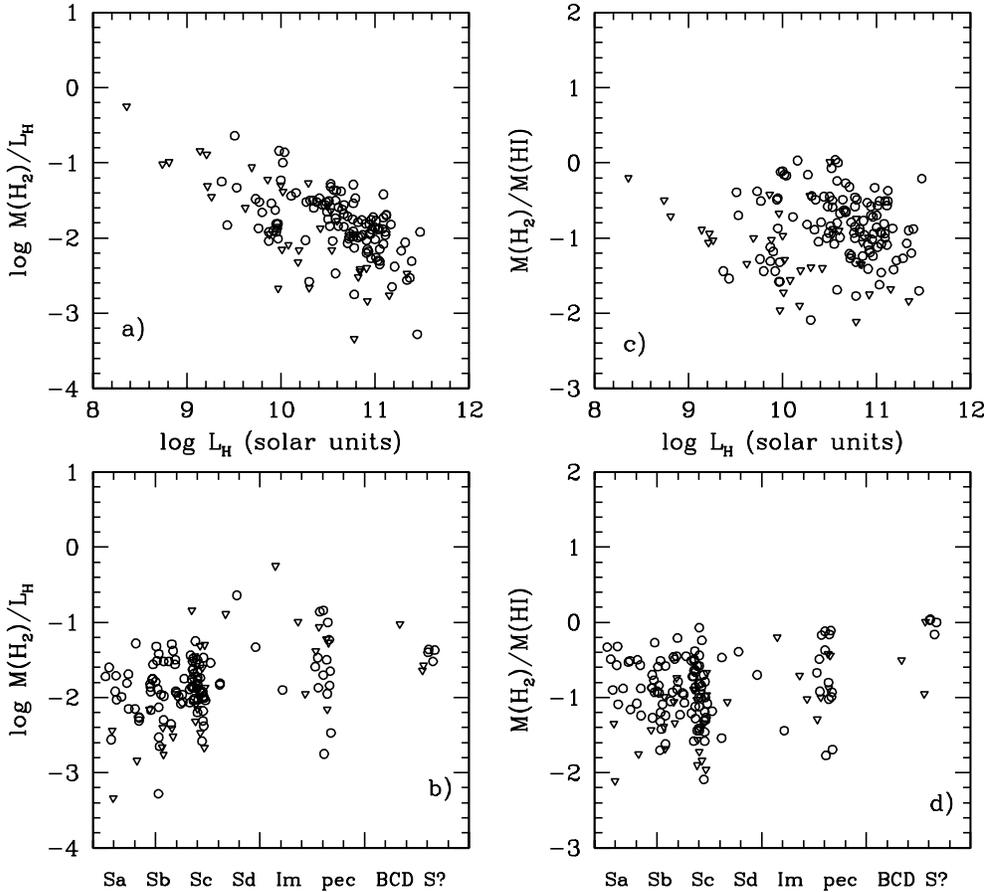}
}
\caption{The relationship between the normalised molecular gas mass of unperturbed 
galaxies ($HI-deficiency$ $\leq$0.3) and a) the H band
luminosity, b) the morphological type; the relationship between the H$_2$ to HI
gas ratio  and c) the H band luminosity, 
d) the morphological type. All quantities are in solar units.
Open triangles are upper limits to the molecular gas mass.
To avoid overplotting, a random number between -0.4 and 0.4 was added to
each half Hubble class bin taken as unity.}
\label{fig.6}
\end{figure*}

The molecular gas content of our sample galaxies, normalized to the total mass of galaxies, 
is plotted on Fig. 6a and b as a function of the H luminosity and the morphological type.
To avoid any systematic environmental effect, only unperturbed 
galaxies with an HI-deficiency $\leq$0.3 (defined as in sect. 5.1)
are considered.\\

Figure 6 shows a strong anticorrelation between the normalized molecular gas mass
and the total mass of galaxies, as traced by the H luminosity.
The relationship with the morphological type is significantely weaker, 
even though it seems that early spirals have a lower normalized  amount of molecular 
gas than late-type ones.
Figure 6c and d show that the molecular to atomic hydrogen ratio
is roughly constant in galaxies spanning a large range in luminosity and
morphological type. The average value is $M(H_2)$/$M(HI)$=0.14 
(with upper limits treated as detections).\\

The weak trends of the molecular gas content (per unit mass)
with the morphological type and with the H luminosity as well as the constant
molecular to atomic gas fraction observed by Boselli et al. (1997b) on a small, 
optically selected sample of Sa-Sc galaxies in the Coma supercluster 
is confirmed here with higher statistical significance and extended to
lower luminosities, including Im and BCDs.
The decrease of the molecular hydrogen to dynamical mass ratio or to the atomic hydrogen
ratio claimed by Casoli et al. (1998), Sage (1993), Young \& Knezek (1989)
and Kenney \& Young (1988b) for Scd-Sm-Im galaxies is probably due to a 
systematic underestimate of the total $M(H_2)$ in low mass galaxies when derived 
assuming a constant $X$ conversion factor.

\subsection{The effect of the environment on the molecular gas}

The available data can be used to analyse the effects of environment on the
molecular gas content of normal galaxies. Following Boselli et al. 
(1997b) an H$_2$ deficiency parameter can be defined once the relationship 
between the molecular gas content and the H luminosity is calibrated
on the unperturbed sample:

\begin{equation}
{log M(H_2)=3.28(\pm0.39)+0.51(\pm0.05)logL_H}
\end{equation}

\noindent
where $M(H_2)$ and $L_H$ are expressed in solar units.
We can thus define the H$_2$ deficiency parameter as:

\begin{equation}
{H_2-deficiency=log M(H_2)_e - log M(H_2)_o}
\end{equation}

\noindent
where $M(H_2)_e$ is the expected molecular gas mass of
a galaxy of a given H luminosity as determined from eq. (8) and
$M(H_2)_o$ is the observed molecular gas mass.
The H$_2$ deficiency parameter determined 
for the whole sample is plotted in Fig. 7 versus the HI-deficiency parameter.

\noindent

The present work confirms the lack of molecular-gas deficient galaxies in 
clusters such as Coma (Boselli et al. 1997b, Casoli et al. 1991) or
Virgo (Kenney \& Young 1989; Boselli 1994), extending previous
results to lower luminosities. This analysis 
suggests that the low luminosity, CO deficient spiral galaxies
observed in Virgo by Kenney \& Young (1988b) are not necessarily
deficient in molecular hydrogen.\\

\begin{figure}
\vbox{\null\vskip 8. cm
\includegraphics{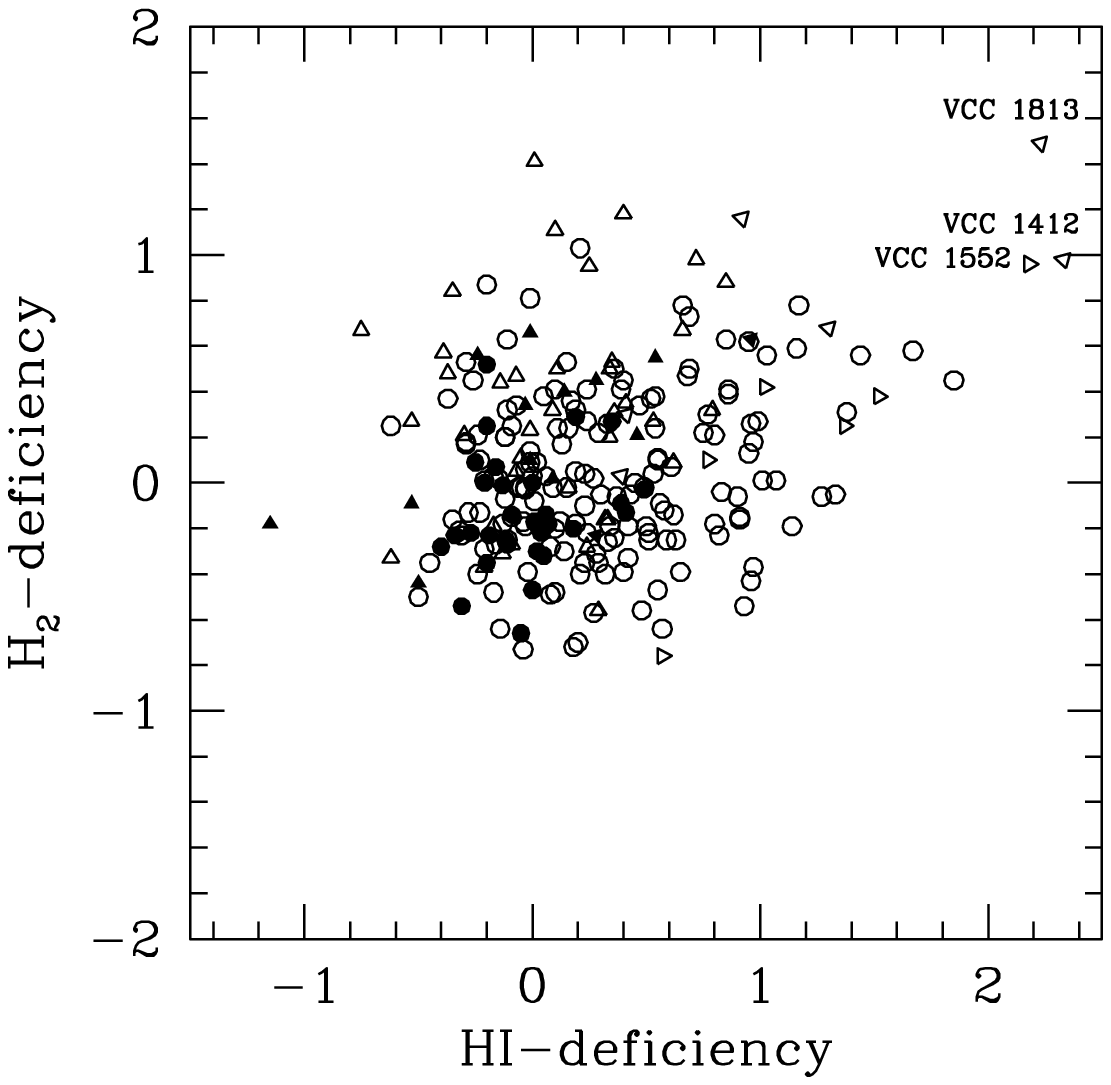}
}
\caption{The relationship between the $H_2$ and the $HI-deficiency$ parameter;
filled dots are for the isolated galaxies sample. Triangles indicate lower 
limits to the H$_2$ ($\triangle$), HI ({\large {$\triangleright$}}) or both gas deficiency.
}
\label{fig.7}
\end{figure}

\subsection{The relationship between the molecular gas content and star formation}

The atomic gas has to condense into molecular clouds to form new stars.
A strong relationship between any tracer of star formation and the molecular
gas content of late-type galaxies is thus expected. 

\begin{figure*}
\vbox{\null\vskip 8. cm
\includegraphics{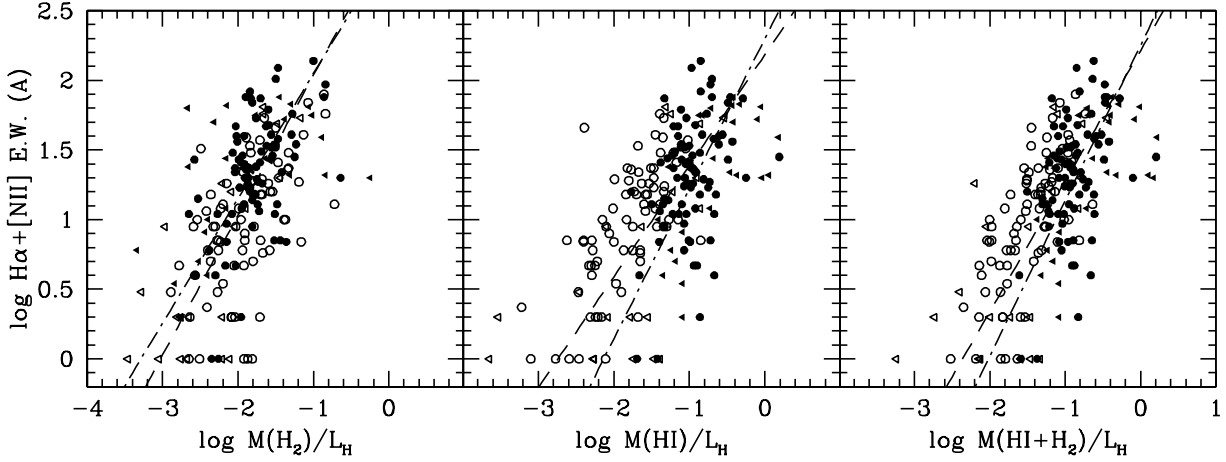}
}
\caption{The relationship between the normalized star formation 
index $H\alpha+[NII] E.W.$ (see text)
and a) the molecular, b) the atomic and c) the total gas mass normalized to the H 
luminosity (in solar units). Filled symbols are for 
the unperturbed sample ($HI-deficiency$ $\leq$ 0.3). Triangles indicate upper 
limits to the molecular hydrogen mass. The dashed line is the best fit to the
whole sample, the dashed-dotted line the best fit to the unperturbed sample.}
\label{fig.8}
\end{figure*}

\begin{table}
\caption[]{
The relationships between the $log H\alpha+[NII] E.W.$ and the atomic, molecular
and total gas phases for the whole sample and for the unperturbed sample (bisector fit,
with upper limits treated as detections).}

Whole sample
\begin{flushleft}
\begin{tabular}{lccc}
\hline
Variable & slope & constant & scatter\\
\hline
$log M(HI)/L_H $                &  0.797  & 2.184 & 0.603\\ 
$log M(H_2)/L_H$                &  1.005  & 3.042 & 0.723\\
$log M(HI+H_2)/L_H$             &  0.938  & 2.214 & 0.554\\
\hline
\end{tabular} 
\end{flushleft}
Unperturbed sample ($HI-def$ $\leq$0.3)
\begin{flushleft}
\begin{tabular}{lccc}
\hline
Variable & slope & constant & scatter \\
\hline
$log M(HI)/L_H $                &  1.078  & 2.300 & 0.902\\ 
$log M(H_2)/L_H$                &  0.896  & 2.944 & 0.741\\
$log M(HI+H_2)/L_H$             &  1.116  & 2.251 & 0.792\\
\hline
\end{tabular} 
\end{flushleft}
\end{table}

\noindent
The relationship between the normalized star formation index $H\alpha+[NII] E.W.$
\footnote{The $H\alpha+[NII] E.W.$ is a normalized entity since it is
defined as the H$\alpha$+[NII] flux normalized to the underlying
red continuum.}
and the molecular gas content (per unit mass) observed in bright galaxies 
by Boselli et al. (1995b, 1997b) extends to low luminosity galaxies
(8 $\leq$ $log L_H$ $\leq$ 12 L$_{H\odot}$) (see Fig. 8a) 
once their molecular gas mass is estimated using a luminosity-dependent 
$X$ conversion factor. The best fits to the data along with the scatter 
from the linear fit are given in Table 6
for the whole sample and for the unperturbed sample ($HI-def$ $\leq$ 0.3).
This observational evidence confirms that the
lack of a strong relationship between any star formation tracer and
the molecular gas mass when determined assuming a constant value of $X$
is due to a systematic underestimate of the total molecular gas mass
of low luminosity galaxies. Figure 8 shows that the relationship 
between the star formation activity and the total
gas (HI $+$ H$_2$) content of galaxies is stronger and less dispersed than for the 
individual gas components. This relationship is shared by normal and gas deficient 
galaxies, even though unperturbed galaxies have on average higher values of
$H\alpha+[NII] E.W.$

\subsection{The star formation efficiency}

The efficiency in transforming gas into stars can be estimated using
the star formation rate (SFR) according to Boselli et al. (2001), i.e.
using H$\alpha$+[NII] fluxes corrected for the contribution 
of the [NII] line, for extinction, and transformed into SFR (in solar masses per year)
assuming a IMF of slope $\alpha$=-2.5 in the mass range $M_{up}$=80 \rm{M$_{\odot}$} 
and $M_{low}$=0.1 \rm{M$_{\odot}$}.\\

\begin{figure*}
\vbox{\null\vskip 7. cm
\includegraphics{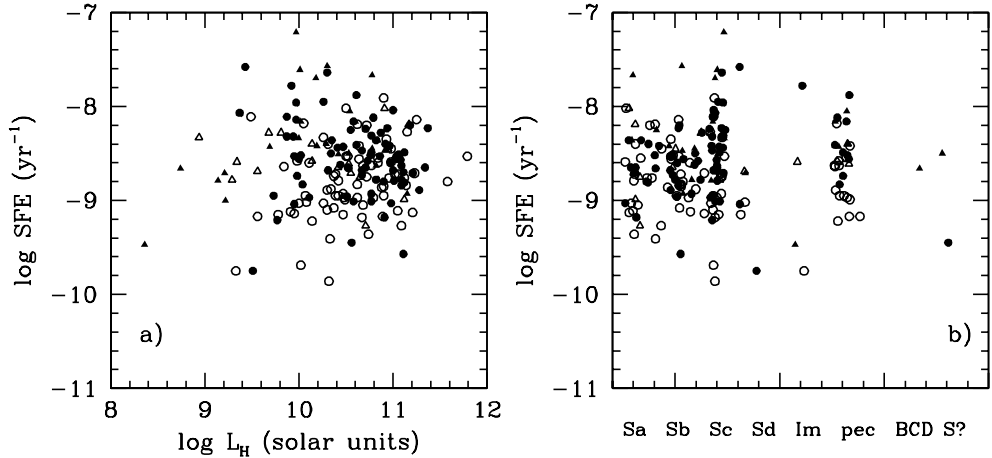}
}

\caption{The relationship between the present star formation efficiency SFE and a) the H luminosity and b)
the morphological type. Filled symbols are for 
the unperturbed sample ($HI-deficiency$ $\leq$ 0.3). Triangles indicate lower 
limits to the SFE.}
\label{fig.9}
\end{figure*}

\noindent
Figure 9 shows the relationship between the present SFE, defined as:

\begin{equation}
{SFE = SFR/M(H_2)~~~~~~  ({\rm yr^{-1}})}
\end{equation}

\noindent
and the H luminosity or the morphological type for the sample galaxies.
If we were to consider the SFE on cosmological timescales,
the whole gas reservoir (HI $+$ H$_2$) would have to be taken into account,
as done in Boselli et al. (2001). 
Figure 9 shows that the present SFE is similar 
for galaxies of different morphological type or luminosity. This result does
not change if the SFE is determined using the total gas mass 
(Boselli et al. 2001; Boissier et al. 2001), as expected being the HI to H$_2$
ratio constant (Fig. 6). Furthermore we do not see
significant systematic differences in the SFE, as determined
using H$_2$ gas masses (eq. 10), for the HI-deficient objects.  

\section{Summary and conclusion}

Using spectro-photometric data available in the literature for a small 
sample of nearby galaxies we have analysed the relationships between the  
$X$ = $N(H_2)$/$I(CO)$ conversion factor and various parameters
characterizing the physical properties of the ISM. 
The behaviour of the nearby galaxies, whose value of $X$ has been 
measured by independent techniques,
is compared to that of a sample of 266 late-type galaxies with available
multifrequency data. For these 266 galaxies $X$ is estimated  
by assuming a metallicity-dependent dust to gas ratio, where the 
dust masses have been determined using ISOPHOT and IRAS 
100-200 $\mu$m data. 
Both samples show an anticorrelation (correlation) between $X$ and the
metallicity (or the UV radiation field), with 
higher values of $X$ in metal-poor, star forming galaxies.
These observational results can be explained if the diffuse
carbon monoxide in the outskirts of the giant molecular clouds is dissociated by 
UV photons at a rate higher than the molecular hydrogen which, given
its higher density, is more efficiently self-shielded.
The ratio between $N(H_2)$ and $I(CO)$ is thus expected to increase
when the UV radiation field increases, as in star forming 
regions or in regions where the extinction, therefore the dust
content and the metallicity, are low. \\
Given the strong relationship between metallicity, star formation
activity and luminosity, the sample of nearby galaxies is used to 
calibrate a luminosity-dependent $X$ conversion factor.
Using this new calibration, we re-analyse the molecular
gas properties of galaxies spanning a large range in morphological
type and luminosity.
Low-mass, dwarf galaxies have higher molecular gas masses (per 
unit galaxy mass) than early-type, massive spirals. The molecular gas
fraction in clouds or complexes is $\sim$ 15 \% of the total HI reservoir for all 
late-type galaxies.\\
Galaxies strongly interacting with the cluster environment have, 
on average, a molecular gas content comparable to isolated, unperturbed 
objects. \\
The star formation rate of late-type galaxies is strongly 
related to their molecular gas content. This relationship is valid for
galaxies spanning a large range in luminosity.\\
The efficiency in transforming gas into stars is roughly constant 
in galaxies of different type and luminosity and belonging to
different environments.

\acknowledgements

We want to thank the 12m telescope operators for their invaluable help
during the remote observations. We thank J. Kenney for providing us with
some unpublished CO data, and C. Hoopes and N. Devereux for the $H\alpha+[NII] E.W.$
of NGC 891 and M33.
We thank C. Bonfanti for providing us with some metallicity measurements,
V. Buat, A. Contursi and J.M. Deharveng for interesting discussions, and
S. Zibetti for providing us with fitting routines.
We thank the anonymous referee for comments and suggestions which helped improving the quality of the manuscript.
This publication makes use of data products from the Two Micron All Sky Survey, 
which is a joint project of the University of Massachusetts and the
Infrared Processing and Analysis Center, funded by the National Aeronautics 
and Space Administration and the National Science Foundation.

\end{document}